\documentclass[amsmath,amssymb,aps,pra,superscriptaddress,twocolumn,floatfix]{revtex4-2}

\usepackage{graphicx}
\usepackage{amsmath}
\usepackage{amsthm}
\usepackage{amssymb}
\usepackage{hyperref}
\usepackage[utf8]{inputenc}
\usepackage{mathtools}
\usepackage[english]{babel}
\usepackage{bbm}
\hypersetup{colorlinks=true, linkcolor=blue, citecolor=blue, urlcolor=blue}
\usepackage{xcolor}
\usepackage[normalem]{ulem}
\usepackage{braket}

\newcommand{\ts}{\textsuperscript}

\newcommand{\PT}{\mathcal{PT}}
\newcommand{\Heff}{H_{\mathrm{eff}}}
\renewcommand{\k}{\mathbf{k}}
\newcommand{\p}{\mathbf{p}}
\newcommand{\E}{\mathbf{E}}
\renewcommand{\r}{\mathbf{r}}
\newcommand{\rhat}{\mathbf{\hat{r}}}
\newcommand{\bwp}{\pmb{\wp}}
\newcommand{\phat}{\mathbf{\hat{p}}}

\newcommand{\lsquare}{\left[}
\newcommand{\rsquare}{\right]}

\newcommand{\abs}[1]{\lvert #1 \rvert} 
\newcommand{\rcvect}[1]{
  \begin{pmatrix}#1\end{pmatrix}%
}

\DeclareMathOperator{\Tr}{Tr}

\begin{document}




\title{Examining the quantum signatures of optimal excitation energy transfer}

\author{Jonah S. Peter}
\email{jonahpeter@g.harvard.edu}
\affiliation{Physics Department, Harvard University, Cambridge, MA 02138}
\affiliation{Biophysics Program, Harvard University, Boston, Massachusetts 02115, USA}
\author{Raphael Holzinger}
\affiliation{Institute for Theoretical Physics, University of Innsbruck, Technikerstrasse 21a, A-6020 Innsbruck, Austria}
\author{Stefan Ostermann}
\affiliation{Physics Department, Harvard University, Cambridge, MA 02138}
\author{Susanne F. Yelin}
\affiliation{Physics Department, Harvard University, Cambridge, MA 02138}

\begin{abstract}
The transport and capture of photo-induced electronic excitations is of fundamental interest to the design of energy efficient quantum technologies and to the study of potential quantum effects in biology. Using a paradigmatic quantum optical model, we examine the influence of coherence, entanglement, and cooperative dissipation on the transport and capture of excitation energy. We demonstrate that the rate of energy extraction is optimized under conditions that minimize the quantum coherence and entanglement of the system, which is a consequence of spontaneous parity time-reversal symmetry breaking. We then examine the effects of vibrational disorder and show that dephasing can be used to enhance the transport of delocalized excitations in settings relevant to biological photosynthesis. Our results highlight the rich, emergent behavior associated with the quantum-to-classical transition with relevance to the design of room-temperature quantum devices.
\end{abstract}

\maketitle

\section{Introduction\label{sect:intro}}

The efficient transport of photo-induced electronic excitations is an important process mediating long-range energy transfer in atomic, molecular, and artificial quantum systems. Transport in isolated systems (i.e., ``closed" quantum systems) is driven by the build-up of quantum coherences that mediate interactions between quantum emitters at spatially separated sites. As compared to classical transport, the quantum properties afforded by these coherences---such as entanglement---can allow for exponential speedups in information transfer~\cite{childs_example_2002} and may be useful for designing new quantum technologies~\cite{higgins_superabsorption_2014}. In practice, however, these processes are invariably subject to environmentally-induced decoherence that destroys the well-defined phase relationships required for quantum superpositions~\cite{nelson_role_2018, schlosshauer_decoherence_2005, schlosshauer_decoherence_2007, zurek_decoherence_2003}. The coupling of such ``open" quantum systems to lossy external environments therefore limits any possible advantages provided by quantum coherence or entanglement. When the system-environment couplings are large, the transport process admits an effectively classical description given by a series of incoherent rate equations. As such, the traditional approach to designing new quantum devices has been to engineer platforms in which these environmental couplings are minimized, with the express purpose of preserving quantum correlations over long timescales. 

However despite the role of decoherence in limiting the quantum properties of excitation transport, recent studies have shown that, in some cases, environmental fluctuations can actually improve the rate of energy transfer. The crucial role of dephasing in overcoming Anderson localization \cite{anderson_absence_1958} in disordered spin networks \cite{maier_environment-assisted_2019}, molecular aggregates \cite{rebentrost_environment-assisted_2009, chin_noise-assisted_2010, plenio_dephasing-assisted_2008, mohseni_environment-assisted_2008, rebentrost_role_2009} and quantum dot arrays \cite{contreras-pulido_dephasing-assisted_2014, contreras-pulido_coherent_2017} is now an active area of research. The crossover between coherent and incoherent dynamics has also attracted significant interest in the study of biological photosynthetic complexes where the role of long-lived coherent oscillations remains controversial~\cite{engel_evidence_2007, ishizaki_quantum_2010, ishizaki_quantum_2012, scholes_using_2017, huelga_vibrations_2013, zerah_harush_photosynthetic_2021, cao_quantum_2020, duan_nature_2017}. Further analyses of quantum signatures in the presence of decoherence are therefore fundamental to the characterization of energy transfer processes in artificial and biological systems alike. A more complete understanding of mixed quantum-classical interactions in both the transport and extraction (or ``trapping") of excitation energy may also lead to the design of novel biomimetic light-harvesting technologies that operate at room temperature~\cite{mattioni_design_2021}.

The underlying task of how to describe, overcome, and ultimately leverage system-bath interactions to efficiently extract energy from a quantum optical system motivates a more in-depth exploration of the quantum-to-classical transition in realistic electromagnetic environments. Within the context of biological and artificial light-harvesting, the inter-emitter interactions are typically assumed to be those of near-field electrostatic dipoles, and the bath dynamics are often limited to models of independent decay and/or pure dephasing~\cite{jang_delocalized_2018, cao_optimization_2009, rebentrost_environment-assisted_2009, chan_single-photon_2018, plenio_dephasing-assisted_2008, mohseni_environment-assisted_2008, rebentrost_role_2009}. These simplifications are warranted only for localized excitations at very short time scales, or when vibrational fluctuations render phase coherences negligible. However, the full light-matter interaction between \emph{quantum} electric dipoles also involves cooperative dissipation that leads to the formation of delocalized superradiant and subradiant modes. These collective modes are essential to the accurate description of quantum dynamics in excitation transport systems~\cite{lehmberg_radiation_1970, lehmberg_radiation_1970-1, asenjo-garcia_exponential_2017, reitz_cooperative_2022, peter_chirality_2024, holzinger_harnessing_2024, peter_chirality-induced_2024} and can greatly influence how these systems respond to environmental decoherence. Consideration of these cooperative effects within the context of biologically-inspired design may also offer new insights into optimal quantum device engineering.

In this work, we seek to elucidate the fundamental mechanisms that drive efficient energy transfer between quantum emitters---with particular emphasis on the role of quantum coherence and entanglement. Using a paradigmatic model of excitation transport and trapping, we present a series of general results linking a wide variety of optimal transport conditions to the quantum-to-classical transition. In particular, we demonstrate that, counterintuitively, optimal transport occurs for excitation trapping rates that minimize the total entanglement of the system. We show that this finding is not limited to disordered or high temperature systems but is instead a fundamental consequence of spontaneous parity time-reversal ($\PT$) symmetry breaking associated with the energy extraction process. We then highlight the influence of cooperative decay and compare the effects of both vibrational and static disorder on localized and delocalized exciton states. Our results demonstrate that dephasing can be used to enhance the transport of delocalized excitations in settings relevant to biological photosynthesis.

The remainder of the manuscript is organized as follows. Section~\ref{sect:model} introduces the cooperative light-matter interactions relevant for long-range quantum transport and excitation trapping. Section~\ref{sect:long-range} discusses how minimizing quantum coherence and entanglement via $\PT$ symmetry breaking leads to the optimal extraction of excitation energy. Sections~\ref{sect:two-site} and ~\ref{sect:nn} relate these results to exceptional points in non-Hermitian optics. Section~\ref{sect:DAT} demonstrates how vibrational-induced dephasing can enhance transport by disrupting bright state emission in settings relevant to biological photosynthesis. Section~\ref{sect:static} discusses how level broadening can enhance the robustness of excitation trapping in the presence of static disorder. Section~\ref{sect:conclusion} summarizes the main results and offers perspectives on future work.

\section{Theoretical model\label{sect:model}}

\begin{figure}
\centering 
\renewcommand\figurename{FIG.}
\includegraphics[width=\textwidth/2]{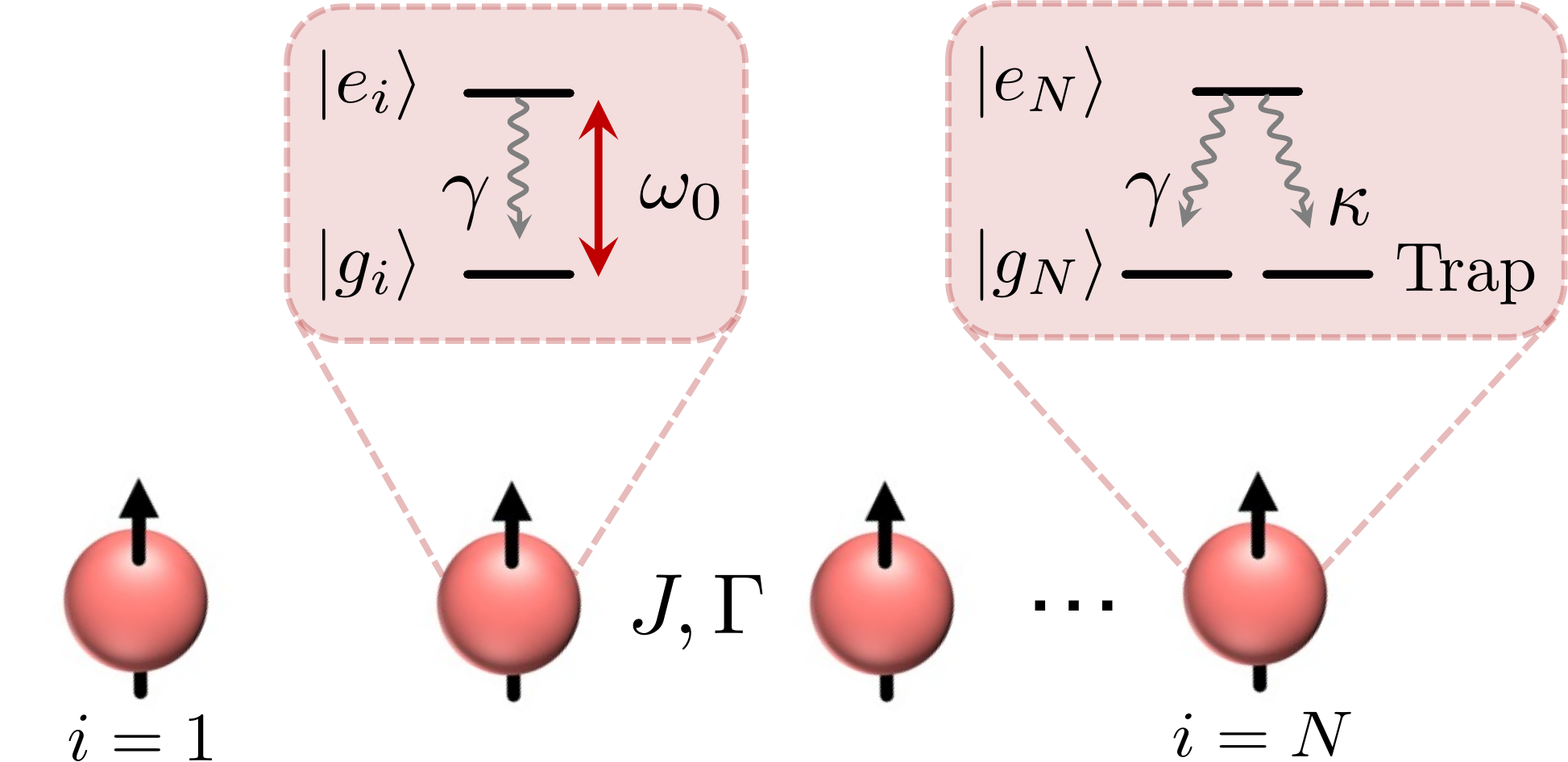}
\caption{Schematic drawing of the excitation transport-trapping process in a one-dimensional chain of quantum emitters. Excitations within the chain are transported between emitters via resonant dipole-dipole interactions $J_{ij}$. Each emitter also experiences independent spontaneous emission with rate $\gamma$ and cooperative radiative decay with rate $\Gamma_{ij}$. The ``acceptor" emitter at site $i=N$ is incoherently coupled to an external trap state with rate $\kappa$, leading to the irreversible extraction of energy from the system.}
\label{fig:schematic}
\end{figure}

We are interested in describing the quantum transport and subsequent trapping of photo-induced electronic excitations between atoms or atom-like emitters (e.g., chromophore molecules, quantum dots). In a complete description, this transport is mediated by interactions with a quantized radiation field and includes cooperative effects that result from collective coupling to the vacuum field modes. In later sections, we also examine the influence of additional environmental fluctuations that induce both dynamical (dephasing) and static (frequency shifts) disorder. For clarity of presentation, we focus here on the key concepts required to understand the results of the main text. Additional details and derivations can be found in the appendices. 

We consider the paradigmatic example of a one-dimensional chain of $N$ equally spaced two-level emitters, each with an excited state $\ket{e_i}$ and ground state $\ket{g_i}$ separated by a resonance frequency $\omega_0 = 2\pi c / \lambda_0$, where $c$ is the speed of light and $\lambda_0$ is the wavelength of the electronic transition (Fig~\ref{fig:schematic}). Coherent excitation transport between the emitters is described by the Hamiltonian
\begin{equation}\label{eq:H_coh}
    H = \sum_{i=1}^N \omega_{0} \sigma_i^\dag \sigma_i + \sum_{i\neq j} J_{ij} \sigma_i^\dag \sigma_j,
\end{equation}
where the transition operators $\sigma_i^\dag = \ket{e_i}\bra{g_i}$ and $\sigma_i = \ket{g_i}\bra{e_i}$ describe the raising and lowering of an excitation at site $i$. The emitters are coupled to one another via resonant dipole-dipole interactions with rate
\begin{multline}\label{eq:J}
    J_{ij} =  \frac{3 \gamma}{4} \left\{ [3(\phat_i \cdot \rhat)(\phat_j \cdot \rhat) - \phat_i \cdot \phat_j]  \left(\frac{\cos{\xi}}{\xi^3} + \frac{\sin{\xi}}{\xi^2} \right) \right. \\
    \left. - [(\phat_i \cdot \rhat)(\phat_j \cdot \rhat) - \phat_i \cdot \phat_j] \left(\frac{\cos{\xi}}{\xi} \right) \right \}
\end{multline}
for dipole matrix element vector $\bwp_i = \wp_i \phat$ and relative coordinate $\xi = \omega_0 r/c$, where $\mathbf{r} = r \rhat$ and $r = \abs{\r_i - \r_j}$ (Appendix~\ref{app:light-matter}). The quantity $\gamma$ denotes the independent radiative decay rate of each emitter. Crucially, the dipole-dipole interaction also includes cooperative decay that leads to superradiant and subradiant emission at subwavelength distances~\cite{dicke_coherence_1954, gross_superradiance_1982}. In the limit where the chain admits (at most) a single excitation at any given time, these radiative processes can be described through the anti-Hermitian term
\begin{equation}
    H_R = - \frac{i}{2} \sum^N_{i,j=1} \Gamma_{ij} \sigma_i^\dag \sigma_j,
\end{equation}
where the diagonal elements of the collective emission rate
\begin{multline}\label{eq:G}
    \Gamma_{ij} =  \frac{3 \gamma}{2} \left\{ [3(\phat_i \cdot \rhat)(\phat_j \cdot \rhat) - \phat_i \cdot \phat_j]  \left(\frac{\sin{\xi}}{\xi^3} - \frac{\cos{\xi}}{\xi^2} \right) \right. \\
    \left. - [(\phat_i \cdot \rhat)(\phat_j \cdot \rhat) - \phat_i \cdot \phat_j] \left(\frac{\sin{\xi}}{\xi} \right) \right \}
\end{multline}
are equal to $\gamma$. Though often neglected in studies of photosynthetic energy transfer, the off-diagonal elements of $\Gamma_{ij}$ are essential to the study of nontrivial quantum dynamics and cannot be neglected even in the limit as $\xi \to 0$ (Appendix~\ref{app:collective}). As such, the full Hamiltonian describing excitation transport through the chain---including both independent and collective radiative emission---is given by $H_\mathrm{ch} = H + H_R$.

In addition to transport, the successful transfer of energy from the site of initial photon absorption to a spatially separated location also involves extracting the excitation from the system. This process is of considerable interest to the study of biological photosynthesis where excitations are funneled to a photochemical reaction center, eventually leading to charge separation~\cite{adolphs_how_2006}. Excitation trapping is also an important process in artificial light-harvesting and could be mediated, for example, by nano-wire structures~\cite{higgins_superabsorption_2014,dorn_using_2011,lu_photocurrent_2009} or semiconductor wells~\cite{agranovich_hybrid_2011,lu_nonradiative_2007}. As is applicable in these systems~\cite{rebentrost_environment-assisted_2009, chin_noise-assisted_2010, plenio_dephasing-assisted_2008, mohseni_environment-assisted_2008, caruso_highly_2009}, we assume that the trapping process is irreversible and can be modeled as an additional Markovian decay channel on the $N\ts{th}$ site (denoted the ``acceptor" site) with rate $\kappa$ by the term
\begin{equation}
    H_T = - \frac{i}{2} \kappa \sigma_N^\dag \sigma_N.
\end{equation}
The complete transport-trapping Hamiltonian is then $\Heff = H_\mathrm{ch} + H_T$. As discussed below, the trapping process itself has a significant influence on the energy transfer efficiency and can be a source of nontrivial quantum dynamics.

\section{Excitation trapping and the quantum-to-classical transition\label{sect:optimal_trap}}

\begin{figure*}
\centering 
\renewcommand\figurename{FIG.}
\includegraphics[width=\textwidth]{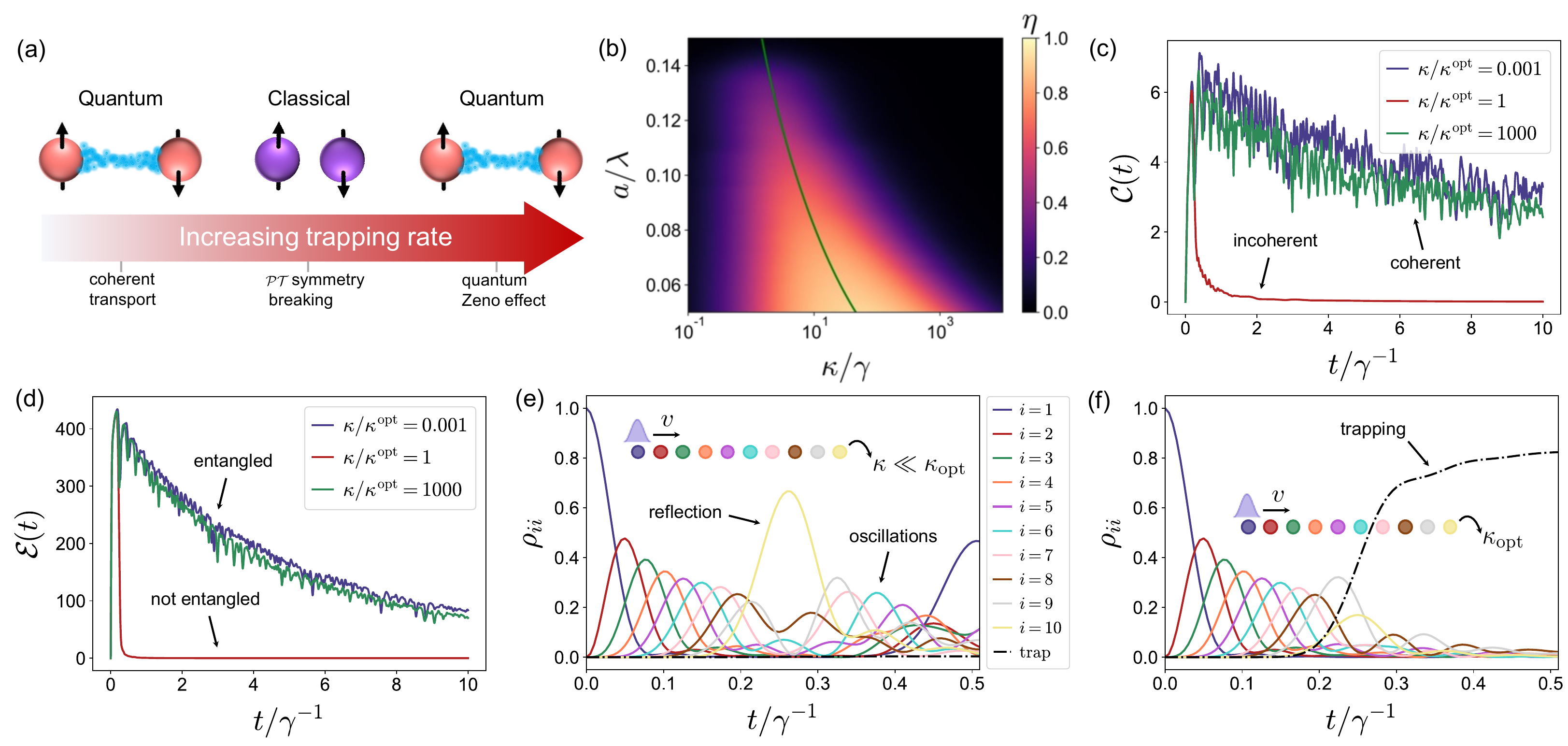}
\caption{(a) Illustration of the quantum-to-classical transition. Small trapping rates lead to coherent oscillations and long-lived quantum entanglement. Optimal excitation trapping occurs for trapping rates that minimize the quantum properties of the system and is accompanied by spontaneous $\PT$ symmetry breaking. At much larger trapping rates, the quantum behavior of the system is restored via the quantum Zeno effect. (b) Transport efficiency $\eta$ for a chain of $N=10$ emitters as a function of the trapping rate $\kappa$ and lattice spacing $a$ at time $\gamma t = 10$. The green line denotes the optimal trapping rate based on the group velocity argument, $\kappa^\mathrm{opt} \approx 2J$. (c) The total quantum coherence $\mathcal{C}(t)$ as a function of time for different values of the trapping rate. (d) The total entanglement $\mathcal{E}(t)$ as a function of time for different values of the trapping rate. (e) Site populations $\rho_{ii}$ for each of the emitters and the trap state when $\kappa / \kappa^\mathrm{opt} = 0.001$. (f) Site populations for $\kappa / \kappa^\mathrm{opt} = 1$. Figure legend is the same as in panel (e). Lattice spacing for panels (c)-(f): $a/\lambda_0 = 0.05$.}
\label{fig:long_range_A}
\end{figure*}

In this section, we show that the optimal trapping condition is one of maximal decoherence and is accompanied by a spontaneous $\PT$ symmetry breaking. We thus demonstrate that varying the trapping rate can facilitate unidirectional energy transfer and can be used to probe the quantum-to-classical transition [Fig.~\ref{fig:long_range_A}(a)]. The main results are presented in Section~\ref{sect:long-range}. Additional supporting analyses based on exceptional point physics are presented in Sections~\ref{sect:two-site} and \ref{sect:nn} using simplified two-site and nearest neighbor models, respectively. 

\subsection{Unidrectional energy transfer through $\PT$ symmetry breaking\label{sect:long-range}}

An initial excitation $\ket{\Psi(0} = \ket{e_1}$ located at one end of the chain will evolve according to $\Heff$ via the Schrodinger equation $i \partial_t \ket{\Psi(t)} = \Heff \ket{\Psi(t)}$. The primary quantity of interest for evaluating the excitation capture process is the transport/trapping efficiency
\begin{equation}
    \eta(\tau) = \kappa \int_0^\tau dt \braket{\Psi(t) | \sigma^\dag_N \sigma_N | \Psi(t)},
\end{equation}
which quantifies the probability that an initial excitation located within the chain at time $t=0$ is found in the trap state at a much later time $t=\tau$. We can study the quantum nature of this process by analyzing the population coherences and entanglement between different emitters. We define the total quantum coherence of the system as the sum of the $l_1$ norms of off-diagonal density matrix elements~\cite{baumgratz_quantifying_2014} $\mathcal{C}(t) = \sum_{i \neq j} \abs{\rho_{ij}(t)}$, where $\rho_{ij} = \braket{\sigma^\dag_i \sigma_j}$. As a measure of quantum entanglement, we consider the logarithmic negativity, which is an entanglement monotone defined for an arbitrary density matrix of a general bipartite system~\cite{plenio_introduction_2007}. The logarithmic negativity for subsystems $A$ and $B$ is given by $E(A|B) = \log_2 \lVert \rho^{T_B} \rVert$, where the superscript $\cdot^{T_B}$ indicates the partial transpose of subsystem $B$ and the operation $\lVert \cdot \rVert$ denotes the trace norm. For the case where there is at most a single excitation in the chain, the logarithmic negativity can be written in terms of the population coherences as~\cite{caruso_entanglement_2010, caruso_highly_2009}
\begin{equation}\label{eq:entanglement}
    E(1,...,k|k+1,...,N) = \log_2(1 - p_0 + \sqrt{p_0^2 + 4X}),
\end{equation}
where $X = \sum_{i=1}^k \sum_{j=k+1}^N \abs{\rho_{ij}}^2$ and $p_0 = 1 - \sum_{i=1}^N \rho_{ii}$. We thus define the total entanglement of the system $\mathcal{E}(t)$ as the sum of the logarithmic negativity across all possible (unique) bipartitions.

Fig.~\ref{fig:long_range_A}(b) shows the trapping efficiency for a chain of $N=10$ transversely polarized emitters as a function of the lattice spacing and the trapping rate. In all configurations, the efficiency exhibits a clear maximum, indicating the existence of an optimal trapping rate $\kappa^\mathrm{opt}$. Strikingly, we find that the trapping process is optimized under conditions that minimize the ``quantumness" of the system. At short times, the inter-site couplings drive the build up of spatial coherences between different emitters. These coherences decay as a consequence of radiative lifetime broadening, but can be long-lived when the coherent couplings are large. As the trapping rate increases, the lifetime of the acceptor site is reduced, and decoherence proceeds more rapidly due to the additional lifetime broadening of the acceptor emitter induced by the trap. Fig.~\ref{fig:long_range_A}(c) shows the time evolution of the total coherence for different values of the trapping rate. At small trapping rates, the coherences undergo damped oscillations with a decay envelope $\propto e^{-t / \tau_\mathcal{C}}$, characterized by a coherence time $\tau_\mathcal{C}$ and an oscillation timescale related to the inter-site coupling strength [see also Eq.~\eqref{eq:TDSE_evo}]. This damping is accompanied by a similar exponential decay of the quantum entanglement with rate constant $\tau_\mathcal{E}$ [Fig.~\ref{fig:long_range_A}(d)]. At the optimal trapping rate, both the total coherence and the total entanglement are critically damped and tend rapidly to zero as the excitation leaves the chain via the trap.

Examination of the site populations $\rho_{ii}$ illustrates this transition from coherent quantum behavior to incoherent classical transport [Figs.~\ref{fig:long_range_A}(e) and (f)]. As the system evolves, the initial excitation acquires a group velocity and proceeds forward to the acceptor site at the opposite end of the chain. When the trapping rate is small, the excitation is not transferred rapidly enough to be captured in its entirety [Fig.~\ref{fig:long_range_A}(e)]. A significant portion is then reflected backwards and can interfere with the remaining forward moving component. This leads to coherent oscillations between the emitters that are damped by the vacuum decay channels. At the optimal trapping rate, the transport becomes unidirectional and proceeds down the chain into the trap without reflection [Fig.~\ref{fig:long_range_A}(f)]. Consequently, the interference phenomena associated with the superposition of the forward and backward moving components is suppressed, and the excitation loses its wave-like nature. 

A prediction of this interpretation is that optimal energy extraction should occur when the rate of transfer to the trap is commensurate with the rate of population arriving at the acceptor via the other emitters in the chain. We can validate this claim by comparing the optimal trapping rate with the group velocities of the resonant chain eigenmodes~\cite{holzinger_harnessing_2024}. In the limit of large $N$ and small dissipation, the eigenenergies of $H_\mathrm{ch}$ form a continuous band and are approximately given by the nearest-neighbor estimate $E(k) \approx 2J \cos{(k a)}$, where $J = J_{i,i+1}$ is the nearest-neighbor hopping rate, $a$ is the distance between nearest-neighbor sites, and $k$ is the lattice quasimomentum. The eigenmodes that are on resonance with the acceptor emitter are located at $ka = \pm \pi/2$ and have group velocity $v = \abs{dE(k)/dk}_{\pi/2} = 2a \abs{J}$. The rate at which population arrives at the acceptor via these resonant modes is then given by $\kappa^\mathrm{opt} = v/a = 2\abs{J}$. This analytical estimate shows excellent agreement with the exact numerical values [Fig.~\ref{fig:long_range_A}(b)].

\begin{figure*}
\centering 
\renewcommand\figurename{FIG.}
\includegraphics[width=\textwidth]{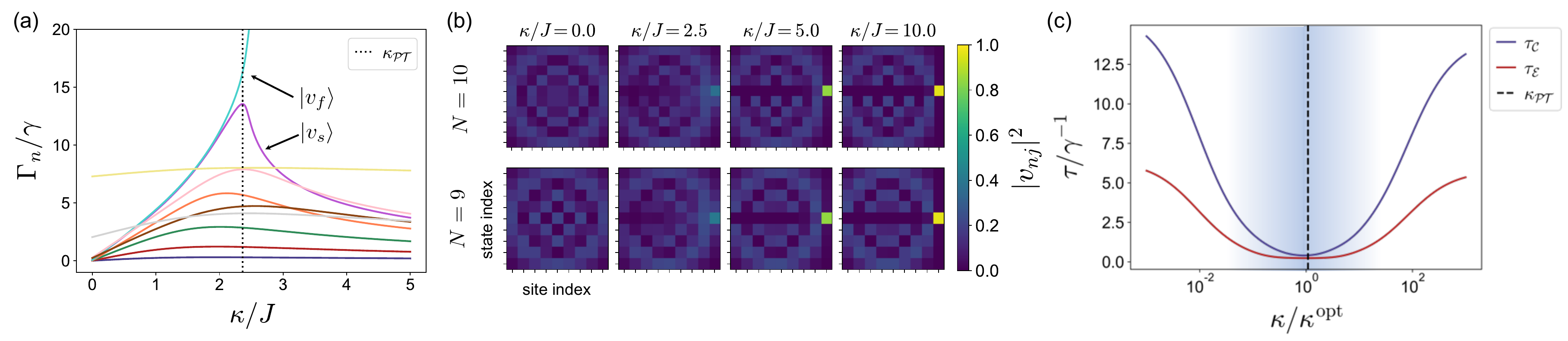}
\caption{(a) Decay rates $\Gamma_n$ for the $N=10$ chain as a function of increasing trapping rate. The system undergoes spontaneous $\PT$ symmetry breaking characterized by a slowly decaying mode with a first derivative that changes sign at $\kappa_\PT$. (b) Excitation probabilities in the site basis for each eigenmode showing progressive localization of the acceptor with increasing trapping rate. (c) Characteristic decay rate of the total system coherence $\tau_\mathcal{C}$ and entanglement $\tau_\mathcal{E}$. Optimal transport occurs near the $\PT$ symmetry breaking transition where both the coherence and entanglement times are minimized. The blue shaded area indicates the classical regime.}
\label{fig:long_range_B}
\end{figure*}

The transition between the coherent (oscillatory) and incoherent (unidirectional) transport regimes is a manifestation of spontaneous $\PT$ symmetry breaking. This symmetry breaking is a known consequence of non-Hermiticity in classical optics~\cite{guo_observation_2009} and can be extended to the quantum optical transport processes described here (see also Sections~\ref{sect:two-site} and \ref{sect:nn} for more detail). Quite generally, $\PT$ symmetry breaking can be identified by examining the eigenmode structure of the non-Hermitian Hamiltonian. The exact eigenvalues of $\Heff$ are complex and given by $\tilde{\varepsilon}_n = \varepsilon_n - (i/2) \Gamma_n$, where $\varepsilon_n$ and $\Gamma_n$ denote the real energy and decay rate of each mode. The onset of $\PT$ symmetry breaking is marked by the existence of an eigenmode with a first derivative in its decay rate that changes sign at the $\PT$ symmetry breaking transition $\kappa_\PT$. We denote this ``slowly decaying" mode as $\ket{v_s}$. That is, $\PT$ symmetry breaking occurs when the decay rate of this mode $\Gamma_s$ satisfies
\begin{equation}\label{eq:passive}
    \left. d\Gamma_s / d\kappa \right \vert_{\kappa = \kappa^-_\PT} > 0 \quad \mathrm{and} \quad \left. d\Gamma_s / d\kappa \right \vert_{\kappa = \kappa^+_\PT} < 0,
\end{equation}
where $\kappa^-_\PT$ and $\kappa^+_\PT$ denote evaluation of the derivative immediately to the left or right of $\kappa_\PT$, respectively~\cite{joglekar_passive_2018, leon-montiel_observation_2018}. Fig.~\ref{fig:long_range_B}(a) shows the behavior of the eigenmode decay rates as a function of the trapping rate. The slowly decaying mode exhibits a sharp decrease in its decay rate near the optimal trapping rate $\kappa^\mathrm{opt}$. Concurrently, the decay rate for the ``fast" mode $\ket{v_f}$ diverges towards $\Gamma_f \to \kappa + \gamma$. The rapid splitting of the fast and slow decay rates is a hallmark feature of spontaneous $\PT$ symmetry breaking. 

As discussed above, the slow trapping regime $\kappa < \kappa^\mathrm{opt} \approx \kappa_\PT$ is dominated by wave-like behavior that results in interference between the incident and reflected excitations. This regime corresponds to the $\PT$ symmetric phase where the oscillatory nature of the quantum transport is most apparent. Trapping rates in the vicinity of $\kappa^\mathrm{opt} \approx \kappa_\PT$ correspond to the $\PT$ broken phase, where the transport is unidirectional down the chain and into the trap. An interesting feature of this system is that the quantum, oscillatory, and $\PT$ symmetric aspects of excitation capture become dominant again at large trapping rates. This is apparent from Figs.~\ref{fig:long_range_A}(c) and \ref{fig:long_range_A}(d), which show the revival of long-lived quantum coherence and entanglement for $\kappa \gg \kappa^\mathrm{opt}$. This phenomenon can be studied in more detail through an eigenmode analysis of $\Heff$. For large $N$ and very small trapping rates ($\kappa \ll J_{ij})$, the collective eigenmodes $\ket{v_n} = \sum_j v_{nj} \ket{e_j}$ are well-approximated by the analytical form \cite{asenjo-garcia_exponential_2017}
\begin{align}\label{eq:ansatz}
    v_{nj} & \approx \sqrt{\frac{2}{N+1}} \cos(k_n x_j), \text{for $n$ odd} \nonumber \\
    v_{nj} & \approx \sqrt{\frac{2}{N+1}} \sin(k_n x_j), \text{for $n$ even},
\end{align}
where $k_n a = \pi n/(N+1)$ and $x_j = j a - (N+1)a/2$. As the trapping rate is increased and the $\PT$ symmetry is broken, the acceptor site becomes isolated from the remaining states. Fig.~\ref{fig:long_range_B}(b) demonstrates this isolation numerically for the $N=10$ chain. Each block shows the site probabilities in the eigenbasis $\abs{v_{nj}}^2$ for different values of the trapping rate. The site index $j$ runs along the columns of each block, whereas the rows denote different eigenstates $\ket{v_n}$. At $\kappa = 0$, the eigenstates are well described by Eq.~\eqref{eq:ansatz}. For $\kappa < \kappa_\PT$, the fast and slow modes $\ket{v_f}$ and $\ket{v_s}$ acquire increasingly larger overlaps with the acceptor site $\ket{e_N}$ with increasing $\kappa$. However, at the $\PT$ symmetry breaking point, $\ket{v_f}$ becomes rapidly localized at the acceptor site and the overlaps with the other sites are drastically reduced. As $\kappa/J \to \infty$, $\ket{v_f} \to \ket{e_N}$ and the remaining eigenstates again become well described by Eq.~\eqref{eq:ansatz} but with $N \to N-1$. In other words, the eigenstates of the $N=10$ chain with strong trapping [top right panel of Fig.~\ref{fig:long_range_B}(b)] are those of the $N=9$ chain at zero trapping [bottom left panel of Fig.~\ref{fig:long_range_B}(b)], plus an additional isolated mode. The original $N$ dimensional Hilbert space $\mathcal{H} = \mathcal{H}_A \oplus \mathcal{H}_B$ is therefore partitioned into two non-interacting subspaces: $\mathcal{H}_A$ consisting of the localized state $\ket{v_f} = \ket{e_N}$, and $\mathcal{H}_B$ spanned by the remaining $N-1$ basis states $\ket{e_i}$ for $i \neq N$. 

The simultaneous decrease in the trapping efficiency that accompanies the isolation of the acceptor site is a direct consequence of the quantum Zeno effect (QZE)~\cite{syassen_strong_2008, garcia-ripoll_dissipation-induced_2009, han_stabilization_2009, zhu_suppressing_2014, cao_optimization_2009, rebentrost_environment-assisted_2009, contreras-pulido_dephasing-assisted_2014, contreras-pulido_coherent_2017}. The high frequency trapping rate acts as a continuous environmental measurement of the acceptor site and prevents transitions to (or from) the remaining, longer-lived states. The suppression of these transitions can be studied by examining the survival probability $S(t)$ for an initial state $\ket{\Psi(0)}$ to be found in the same state at a later time $t$. For an external measurement apparatus, the trend towards $S(t) \to 1$ for increasingly frequent projective measurements is a hallmark of the QZE~\cite{peres_zeno_1980}. Here, we impose only a single projective measurement at time $t$ and show that the high frequency trapping rate leads to a ``freezing" of the coherent time evolution. The survival probability for $\ket{\Psi(0)} = \ket{e_N}$ under Hamiltonian evolution with $\Heff$ is given by 
\begin{align}
    S(t) &= \abs{\braket{e_N | e^{-i\Heff T}| e_N}}^2 \nonumber \\
    &= \abs{\braket{\Psi(0) | \openone - i \Heff T - \frac{1}{2}\Heff^2 T^2 + ...| \Psi(0}}^2,
\end{align}
where the Hamiltonian acts as
\begin{equation}
    \Heff \ket{e_N} = -\frac{i}{2} (\gamma + \kappa) \ket{e_N} + \sum_{i\neq N} (J_{iN} - \frac{i}{2} \Gamma_{iN}) \ket{e_i}
\end{equation}
(for convenience, we subtract off the constant energy $\omega_0$ contribution). In the limit $\kappa \gg J_{iN}, \Gamma_{iN}$, the initial state is an approximate eigenstate of $\Heff$, and the survival probability is $S(t) \approx e^{-(\gamma + \kappa)t}$. The equality becomes exact as $\kappa/J \to \infty$. The exponential decay of $\ket{\Psi(0)}$ is the same as that for an isolated atom (i.e., $J_{iN}, \Gamma_{iN} = 0$). In other words, the initial state does not feel the influence of the surrounding atoms: the coherent evolution is ``frozen in time." 

The effect of this isolation is that initial states that do \emph{not} have overlap with the acceptor site are fully contained within a restricted subspace. In the high frequency trapping limit, the Hamiltonian then takes the form
\begin{equation}
    \Heff \to P \Heff P - \frac{i}{2}(\gamma + \kappa) \sigma^\dag_N \sigma_N
\end{equation}
where the projection operator $P = \sum_{i \neq N} \ket{e_i} \bra{e_i}$ runs over the restricted subspace. A general state on this subspace $\ket{\psi} = \sum_{i\neq N} c_i \ket{e_i}$ is therefore immune to the additional decoherence imposed by the trapping process and cannot escape the chain except by decaying to vacuum. In the limit where $\kappa \gg J_{ij} \gg \gamma, \Gamma_{ij}$, these states form an (approximate) decoherence-free subspace resulting in long-lived quantum coherence and entanglement. The net result is that the QZE effectively restores the $\PT$ symmetry of the system at a cost of reducing the dimensionality of the Hilbert space. Optimal excitation trapping occurs between these two $\PT$ symmetric regimes, where decoherence imposes a directionality on the energy transfer process.

\subsection{Two-site model\label{sect:two-site}}

We can gain some intuition for the results of the preceding section by studying a simple two-site model where the transition from coherent to incoherent evolution and the accompanying $\PT$ symmetry breaking are more apparent. For simplicity, we neglect cooperative dissipation in this discussion ($\Gamma_{ij} = \delta_{ij} \gamma$). In the $\{\ket{e_1}, \ket{e_2}\}$ basis, the non-Hermitian effective Hamiltonian is
\begin{equation}\label{eq:2-atom}
    \Heff = 
    \begin{bmatrix}
        -\frac{i}{2}\gamma & J \\
        J & -\frac{i}{2} (\gamma + \kappa)
    \end{bmatrix}
\end{equation}
(we set $\omega_0 = 0$ and $J \in \mathbb{R}$ without loss of generality). Beginning in the initial state $\ket{\Psi(0)} = \ket{e_1}$, the time evolution of the vector $\ket{\Psi(t)} = c_1(t) \ket{e_1} + c_2(t) \ket{e_2}$ is given by
\begin{align}\label{eq:TDSE_evo}
c_1(t) &= e^{-\chi t/2} \left[\frac{\kappa}{4} \frac{\sin{(\lambda t)}}{\lambda} + \cos{(\lambda t)} \right] \\ \nonumber
c_2(t) &= -iJ e^{-\chi t/2} \frac{\sin{(\lambda t)}}{\lambda},
\end{align}
where we have defined the quantities $\lambda \equiv (1/2) \sqrt{4J^2 - (\kappa/2)^2}$ and $\chi \equiv \gamma + \kappa/2$. It is clear from Eq.~\eqref{eq:TDSE_evo} that both the excited state populations $\rho_{ii}(t) = \abs{c_i(t)}^2$ and the population coherences $\abs{\rho_{ij}(t)} = \abs{c_i(t) c_j^*(t)}$ experience a transition from underdamped to overdamped dynamics when $\kappa = 4J$. The critical damping condition therefore delineates the coherent and incoherent transport regimes. For the Hamiltonian \eqref{eq:2-atom}, this transition is marked by a branch cut singularity at $\lambda=0$, known as an exceptional point (EP). The EP denotes the location where the complex eigenvalues and eigenvectors of the non-Hermitian Hamiltonian coalesce. 

\begin{figure}
\centering 
\renewcommand\figurename{FIG.}
\includegraphics[width=\textwidth/2]{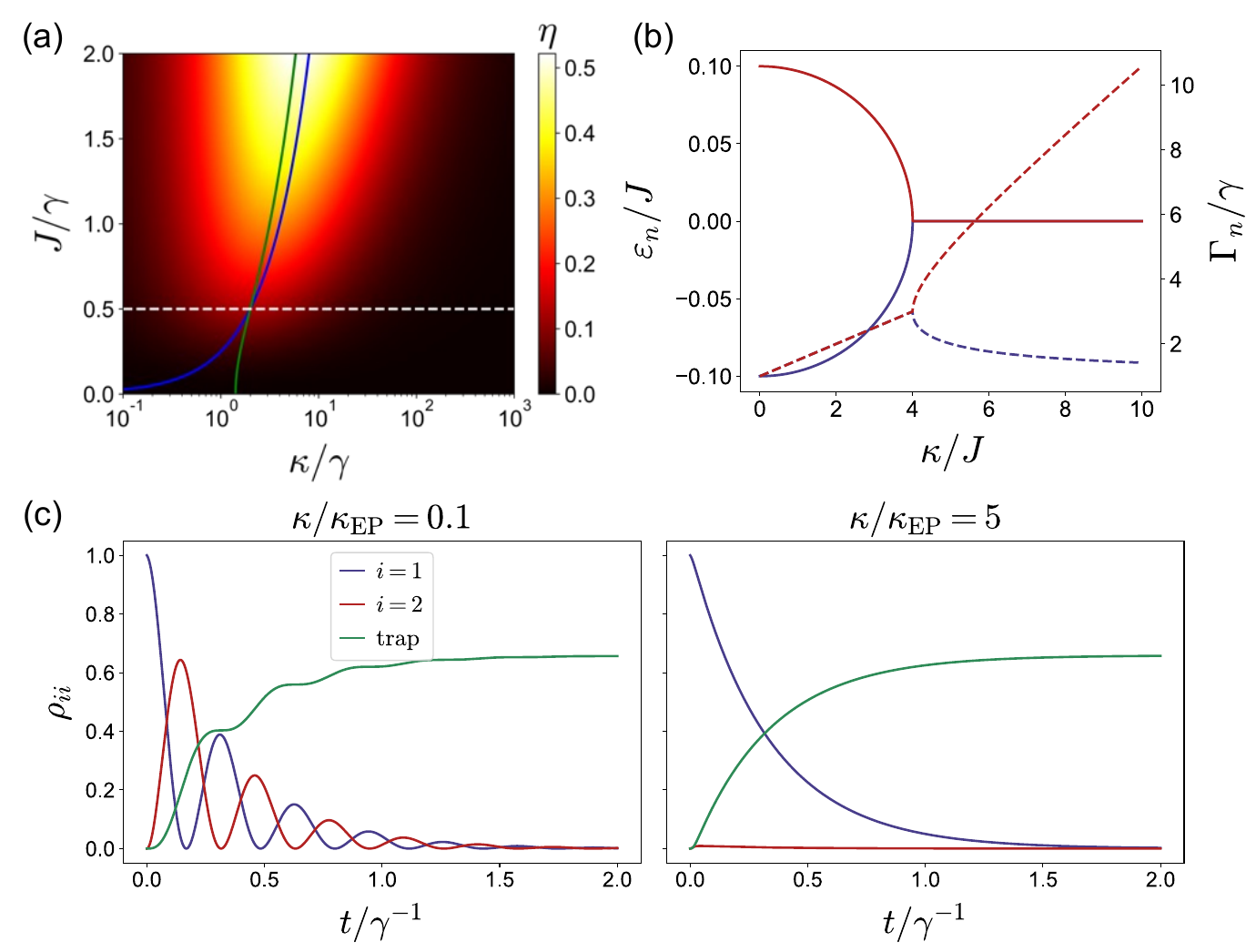}
\caption{(a) Trapping efficiency $\eta(\infty)$ for the two-site model as a function of the trapping rate $\kappa$ and inter-site coupling $J$. The solid green line indicates the optimal trapping rate $\kappa^\mathrm{opt}$ whereas the solid blue line denotes the EP at $\kappa = 4J$. The dashed white line marks where optimal transport coincides with critical damping at $J = \gamma/2$. (b) Energies $\varepsilon_n$ (solid lines) and decay rates $\Gamma_n$ (dashed lines) of the collective eigenmodes showing a branch cut singularity at the EP. (c) Populations of the emitter sites and trap state in the coherent (left) and incoherent (right) trapping regimes.}
\label{fig:2_atom}
\end{figure}

We may further examine the dynamics by solving for the transport efficiency as a function of the trapping rate, $\kappa$. In the limit as $t \to \infty$, the analytical solution is given by 
\begin{equation}
    \eta(\infty) = \frac{4J^2 \kappa}{(2\gamma + \kappa) \lsquare 4J^2 + \gamma(\gamma + \kappa) \rsquare}.
\end{equation}
The efficiency reaches a maximum of 
\begin{equation}
    \eta^\mathrm{opt} = 4J^2 / (4J^2 + 3\gamma^2 + 2\gamma \sqrt{8J^2 + 2\gamma^2})
\end{equation} for the optimal trapping rate $\kappa^\mathrm{opt} = \sqrt{8J^2 + 2\gamma^2}$. At much larger trapping rates, the transport is hindered by the QZE. Fig.~\ref{fig:2_atom}(a) compares the optimal trapping rate with the location of the EP at $\kappa_{\mathrm{EP}} = 4J$. Except at very small coupling strengths ($J \ll \gamma / 2$), the optimal trapping rate is well-approximated by the critical damping condition. As discussed in the previous section, the emergence of irreversible, incoherent dynamics is a manifestation of spontaneous $\mathcal{PT}$ symmetry breaking. The coherent (underdamped) regime corresponds to the $\PT$ symmetric phase where the two eigenmodes differ in energy but have equal decay rates [Fig.~\ref{fig:2_atom}(b)]. In this phase, the $\PT$ operation leaves the eigenmodes invariant, up to a global decay that changes sign (Appendix~\ref{app:PT}). At the EP, the decay rates split into one fast ($\Gamma_f$) and one slow ($\Gamma_s$); the $\PT$ symmetry is spontaneously broken.

Strictly speaking, true $\PT$ symmetry occurs only in balanced gain-loss systems where the Hamiltonian commutes with the $\PT$ operator. The Hamiltonian \eqref{eq:2-atom} is related to a true $\PT$ symmetric Hamiltonian through a global decay term with rate $\chi$,
\begin{equation}\label{eq:shift}
    \Heff = H_\PT - \frac{i}{2} \chi \openone,
\end{equation}
where $[H_\PT, \PT] = 0$ and $\openone$ is the identity operator. The eigenmodes of $\Heff$ and $H_\PT$ are related through a gauge transformation. In this context, $\Heff$ is said to exhibit a ``passive" $\PT$ symmetry that can be defined through properties of its complex eigenvalues, $\tilde{\varepsilon}_n = \varepsilon_n - (i/2)\Gamma_n$. Here, $\varepsilon_n$ and $\Gamma_n$ are the (real) energies and decay rates of each eigenmode. Passive $\PT$ symmetry breaking is then defined by Eq.~\eqref{eq:passive}. For the two-site Hamiltonian, $\kappa_\PT = \kappa_{\mathrm{EP}}$, though the definition based on Eq.~\eqref{eq:passive} remains valid even in the absence of EPs (see Section~\ref{sect:nn} and Ref.~\cite{joglekar_passive_2018}). 


\subsection{Nearest-neighbor model \label{sect:nn}}

In order to study the crossover between the independent decay two-site model (where $\PT$ symmetry breaking occurs at the EP) and the more complicated $N$-site cooperative decay model (where the role of EPs is less clear), we now focus on an $N$-site nearest-neighbor model. We continue to set $\Gamma_{ij} = \delta_{ij} \gamma$, as the inclusion of cooperative decay in a nearest-neighbor transport model can lead to unphysical results. In this case, the nearest-neighbor transport-trapping Hamiltonian is given by
\begin{equation}\label{eq:nn}
    \Heff = \sum_{i=1}^N \left(\omega_0 - \frac{i}{2} \gamma \right) \sigma_i^\dag \sigma_i + J \sum_{\braket{i,j}} \sigma_i^\dag \sigma_j - \frac{i}{2} \kappa \sigma_N^\dag \sigma_N,
\end{equation}
where the angled brackets denote a sum over nearest-neighbor sites. This Hamiltonian also exhibits EPs, which can be found by first solving for the degenerate eigenvalues. After a suitable gauge transformation (see Appendix~\ref{app:EP}), the eigenvalues $\tilde{\varepsilon}$ of Eq.~\eqref{eq:nn} are given by the roots of the characteristic equation
\begin{equation}\label{eq:char_poly}
    \phi_N(x) = J^N \left[ \left(\frac{-i\kappa}{2J} - 2x \right) U_{N-1}(x) - U_{N-2}(x) \right],
\end{equation}
where $x = -\tilde{\varepsilon}/(2J)$ and $U_n(x)$ is the $n$\ts{th} degree Chebyshev polynomial of the second kind. The EPs are then found by setting the discriminant of $\phi_M(x)$ equal to zero and affirming the degeneracy of the associated eigenvectors. It is easy to verify that Eq.~\eqref{eq:char_poly} reproduces the location of the single EP for the two-site model given in the previous section.

\begin{figure}
\centering 
\renewcommand\figurename{FIG.}
\includegraphics[width=\textwidth/2]{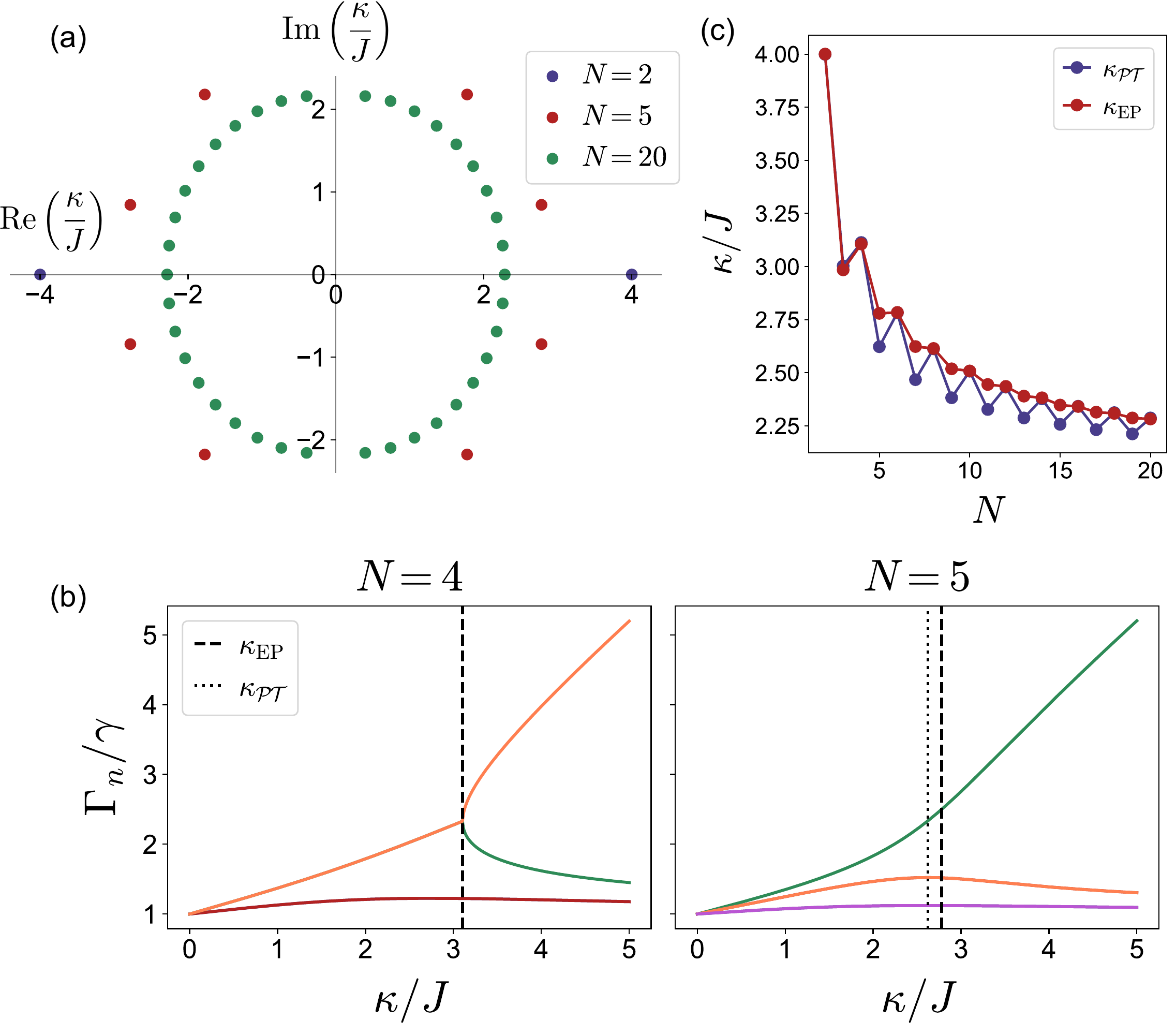}
\caption{(a) Location of the EPs in the complex plain for the nearest-neighbor Hamiltonian. Colors indicate different system sizes: $N=2$ (blue), $N=5$ (red), and $N=20$ (green). (b) Eigenmode decay rates for exemplary even (left) and odd (right) $N$ chains. $\kappa_\mathrm{EP} = \kappa_\PT$ for even $N$. (c) Comparison of the nearest EP projected onto the real axis with the location of the $\PT$ symmetry breaking transition as a function of chain length. The two quantities converge as $N \to \infty$.}
\label{fig:nearest-neighbor.pdf}
\end{figure}

Fig.~\ref{fig:nearest-neighbor.pdf}(a) shows the location of the degenerate eigenvalues in the complex plane for chains of $N=2$, 5, and 20 atoms. There are always $2N - 2$ points where $\phi_N(x)$ has at least one degenerate root. Notably, there is a prominent distinction between systems of even and odd numbers of atoms. Nearest-neighbor chains of even $N$ yield an EP on the positive real axis, corresponding to a physically realizable value of the trapping rate. Similar to the two-site case, this EP is located at a branch cut singularity where the initially identical decay rates of two separate eigenmodes begin to diverge [Fig.~\ref{fig:nearest-neighbor.pdf}(b)]. The lower branch exhibits a first derivative in $\kappa$ that changes sign at the EP, indicating a passive $\PT$ symmetry breaking phase transition. 

For odd $N$, however, all EPs are shifted into the complex plane. The branch cut singularity is therefore inaccessible for real-valued $J$ and $\kappa$, and the eigenmodes never coalesce by varying the trapping rate. Yet despite the obfuscation of the EP, odd $N$ chains continue to demonstrate spontaneous $\PT$ symmetry breaking. The fast and slow modes involved in the EP form an avoided crossing on the real axis as a function of $\kappa$, with the extremum in the slowly decaying branch $d\Gamma_s/d\kappa = 0$ denoting the phase transition at $\kappa = \kappa_\PT$. As for the even $N$ chains, the transition yields two distinct phases: a passive $\PT$ symmetric phase in which the decay rates increase with $\kappa$, and a passive $\PT$ broken phase characterized by a decay rate with a first derivative that changes sign. 

For large $N$, the location of the passive $\PT$ transition is well-approximated by the projection of the nearest EP onto the positive real axis [Fig.~\ref{fig:nearest-neighbor.pdf}(b)]. The two quantities are identical for even $N$. As $N$ increases, the distance between the nearest EP and the real axis decreases (for odd $N$), and the difference between even and odd---that is, between $N$ and $N+1$---tends to zero. One should note that For $N > 2$, Eq.~\eqref{eq:nn} is not isomorphic with any true $\PT$ symmetric Hamiltonian and cannot be identity-shifted in the way of Eq.~\eqref{eq:shift} to form a balanced gain-loss system. Nevertheless, the eigenmodes of the multi-site model still exhibit the same symmetry breaking characteristics of a passive $\PT$ phase transition.

\section{Enhancing transport through vibrational and static disorder}

The previous section illustrates that the directionality imposed by spontaneous $\PT$ symmetry breaking can enhance the efficiency of excitation trapping through the critical damping of spatial coherences. In this section, we look at two other mechanisms of decoherence: vibrational disorder through dynamical dephasing (Section~\ref{sect:DAT}) and static disorder through frequency fluctuations (Section~\ref{sect:static}). We show that proper consideration of cooperative dissipation [Eq.~\eqref{eq:G}] reveals new mechanisms for enhancing long-range excitation energy transfer in biologically-relevant settings, and that level broadening can increase the robustness of the trapping process.

\subsection{Dephasing assisted transport without disorder\label{sect:DAT}}

\begin{figure*}
\centering  
\renewcommand\figurename{FIG.}
\includegraphics[width=\textwidth]{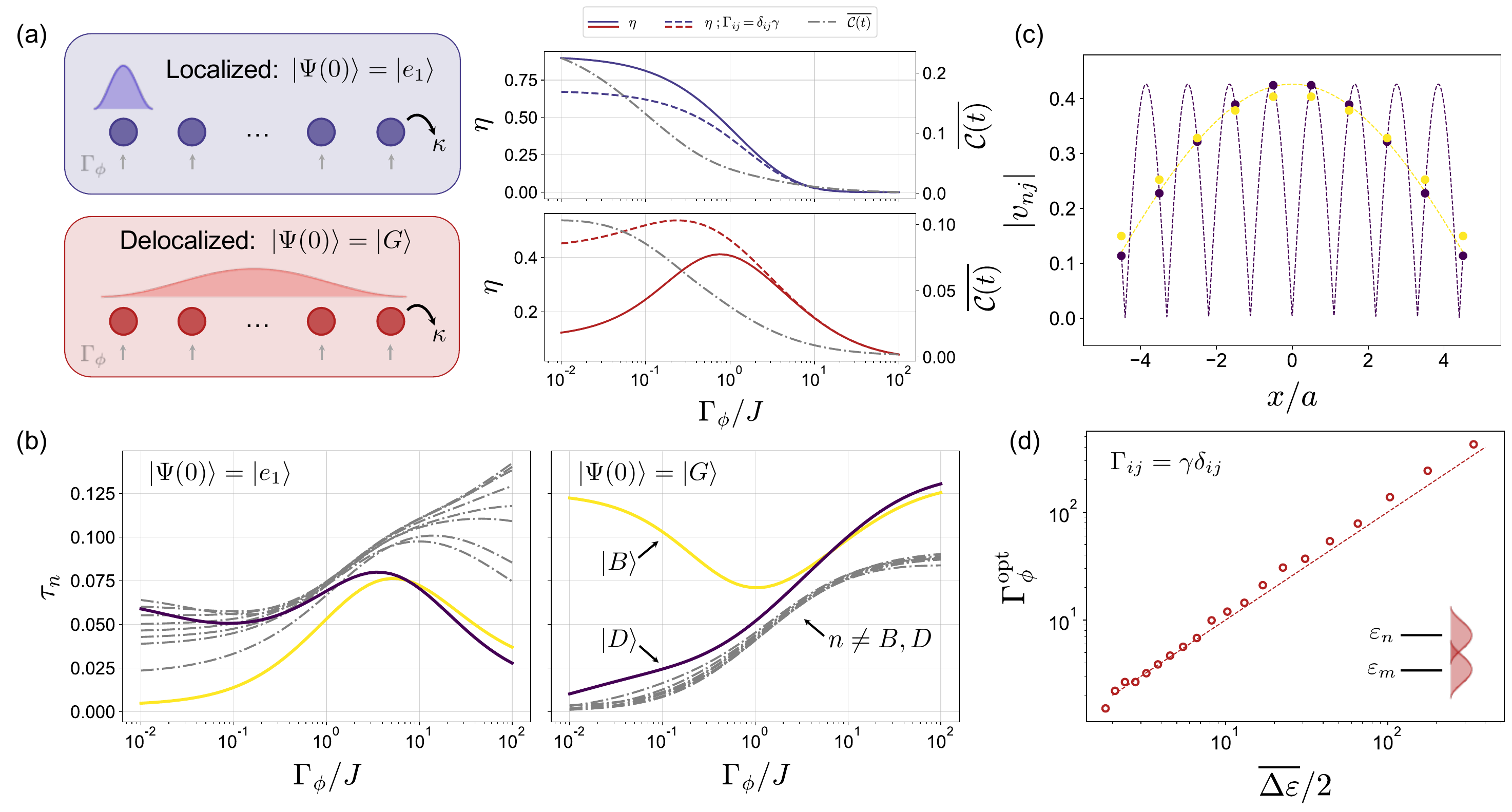}
\caption{(a) Comparison of the transport efficiency between the localized initial state $\ket{e_1}$ (blue) and the delocalized initial state $\ket{G}$ (red) as a function of dephasing strength. Dashed lines show the corresponding transport efficiencies in the absence of collective dissipation. The time-averaged quantum coherence $\overline{\mathcal{C}(t)}$ is shown in grey. (b) Comparison of the mean residence times between the localized (left) and delocalized (right) initial states for each chain mode. Blue and yellow curves indicate the dark and bright states $\ket{D}$ and $\ket{B}$, respectively. (c) Blue and yellow circles denote the site amplitudes for states $\ket{D}$ and $\ket{B}$. Dashed lines show the corresponding values according to the mode ansatz in Eq.~\eqref{eq:ansatz}. (d) Open circles denote the optimal dephasing rate as a function of mean eigenmode energy splitting for the delocalized state in the independent decay model. A slope of $1$ on the log-log plot (dashed line) indicates a linear relationship. Parameters for all panels: $N=10, a = 0.05 \lambda_0, \kappa = 4J, t = 10/ \gamma$.}
\label{fig:dephasing}
\end{figure*}

The traditional picture of vibrational-induced decoherence in the context of excitation transport is that local fluctuations reduce the transport efficiency through dynamical dephasing. A simple example is the exponential decay of perfect Rabi oscillations in the presence of a thermal reservoir. However, numerous studies have since demonstrated that high temperature dephasing can actually enhance transport when the transition frequencies of the emitters are nonuniform~\cite{maier_environment-assisted_2019, rebentrost_environment-assisted_2009, chin_noise-assisted_2010, plenio_dephasing-assisted_2008, mohseni_environment-assisted_2008, rebentrost_role_2009, contreras-pulido_dephasing-assisted_2014, contreras-pulido_coherent_2017}. In these instances, system-bath interactions can overcome the effects of Anderson localization induced by the static frequency disorder. Such phenomena have been termed ``environment assisted quantum transport" or ``dephasing assisted transport" and play a crucial role in modern theories of biological photosynthetic light-harvesting. 

Here we demonstrate a strong dephasing assisted transport enhancement that occurs \emph{without} static disorder. The effect is only revealed through a proper inclusion of off-diagonal cooperative decay [Eq.~\eqref{eq:G}]. At finite temperature, the emitters are subject to thermal fluctuations that couple to the electronic degrees of freedom through the interaction term~\cite{haken_exactly_1973, rebentrost_environment-assisted_2009}
\begin{equation}\label{eq:V}
    H_\phi = \sum_{i=1}^N q_i(t) \sigma^\dag_i \sigma_i.
\end{equation}
Here, $q_i(t)$ denotes classical stochastic fluctuations in the resonance frequencies of the emitters due to the effects of the thermal bath (e.g., chromophore molecules vibrating in a protein scaffold). We assume the fluctuations at spatially distinct sites are independent, identical, and Markovian such that $\braket{q_i(t) q_j(t')} = \Gamma_\phi \delta_{ij} \delta(t-t')$ for homogeneous linewidth $\Gamma_\phi$. The Markov approximation is valid in the high temperature limit $k_B T \gg J$ where the vibrational coherence time is much shorter than the characteristic timescale of the emitters~\cite{capek_hakenstroblreineker_1993, leegwater_coherent_1996}. For unbiased Gaussian fluctuations satisfying $\braket{q_i(t)} = 0$, the dynamics generated by Eq.~\eqref{eq:V} are well described by the pure dephasing Lindbladian
\begin{equation}\label{eq:dephasing}
    \mathcal{L}_\phi [\rho] = \sum_{i=1}^N \frac{\Gamma_\phi}{2} \left ( 2 \sigma^\dag_i \sigma_i \rho \sigma^\dag_i \sigma_i - \{\sigma^\dag_i \sigma_i, \rho \} \right ).
\end{equation}
This Lindbladian leads to a suppression of quantum coherence and therefore probes the quantum-to-classical transition. The resulting equations of motion are then given by
\begin{equation}
    \dot{\rho}(t) = -i \left(H_\mathrm{eff}\rho(t) - \rho(t) H_\mathrm{eff}^\dag \right) + \mathcal{L}_\phi[\rho(t)].
\end{equation}
We note that excitation and stimulated emission by thermal photons remain negligible in this regime so long as the assumption $\omega_0 \gg k_B T$ remains valid (e.g., optical transitions at room temperature). 

The effects of dephasing on excitation transport and trapping are strongly influenced by the degree of localization of the initial state. In particular, for subwavelength chromophore complexes, the incident light field will excite a delocalized state with a symmetric phase distribution. Here, we consider the delocalized Gaussian initial state $\ket{G} = (1/A) \sum_j g(x_j) \ket{e_j}$, where $g(x_j) = \exp{\{-x_j^2 / 2 s^2 \}}$, $x_j$ is defined as in Eq.~\eqref{eq:ansatz}, $s$ is the standard deviation, and $A$ is a normalization factor. To maintain the notion of ``transport," we set $s/a = 3 < N/3$ such that the initial overlap with the acceptor site is negligible---though this is not essential, and similar results are obtained for other states with symmetric phase distributions (e.g., plane waves). The bottom panel of Fig.~\ref{fig:dephasing}(a) shows the transport efficiency for the delocalized state as a function of the dephasing strength. The time-averaged quantum coherence in the site basis is given by $\overline{\mathcal{C}(t)} = (1/t) \int_0^t \mathcal{C}(t') dt'$ and is denoted by the dotted-dashed grey line. Remarkably, thermal fluctuations can strongly improve the transport efficiency in the presence of cooperative dissipation (solid red curve). The enhancement is maximized for dephasing rates on the order of the nearest-neighbor coupling, $\Gamma_\phi / J \approx 1$. This is a strikingly different result than that observed in quantum random walks where the onset of dephasing and the associated loss of quantum coherence greatly reduces transport speed~\cite{brun_quantum_2003, childs_example_2002} (though we note that at very large dephasing rates transport is again suppressed, this time by the QZE~\cite{rebentrost_environment-assisted_2009}). Dephasing assisted transport in delocalized symmetric states thus highlights another key instance where decoherence effects can be leveraged to enhance excitation transport and trapping. 

The dynamics are qualitatively different when initializing in a completely localized state. For the case of $\ket{\Psi(0)} = \ket{e_1}$, the trapping efficiency is a monotonically decreasing function of the dephasing rate [Fig.~\ref{fig:dephasing}(a), solid blue line]. In this instance the conventional wisdom holds, and the rapid decay of inter-site coherences leads to a reduction in trapping efficiency. A comparison with the independent decay model (dashed blue curve) demonstrates that an accurate quantitative assessment of excitation transport at small to moderate dephasing is only possible by including cooperative decay.

The mechanism responsible for enhancing the transport of the delocalized initial state is fundamentally distinct from similar dephasing assisted transport phenomena that have been described previously~\cite{maier_environment-assisted_2019, rebentrost_environment-assisted_2009, chin_noise-assisted_2010, plenio_dephasing-assisted_2008, mohseni_environment-assisted_2008, rebentrost_role_2009, contreras-pulido_dephasing-assisted_2014, contreras-pulido_coherent_2017}. In those works, static frequency disorder induces Anderson localization~\cite{anderson_absence_1958} that suppresses the transport of initially localized states by shifting the site energies off resonance from one another. Dynamical dephasing then acts as a diffusive process and allows the excitation to explore other sites via a classical random walk~\cite{rebentrost_environment-assisted_2009}. By contrast, here we describe a system without static disorder, and where the effect is lost for localized initial states (blue curves) and for non-cooperative decay channels (dashed red curve). Instead, the difference in transport dynamics between the localized and delocalized initial states in the cooperative decay model is primarily due to the existence of subradiant and superradiant collective modes. These modes are eigenstates of the chain Hamiltonian $H_\mathrm{ch}$ and are either dark ($\Gamma_D \ll \gamma$) or bright ($\Gamma_B \gg \gamma$) depending on the local phase relationships. The most subradiant mode $\ket{D}$ and the most superradiant mode $\ket{B}$ are given approximately by Eq.~\eqref{eq:ansatz} for $n=N$ and $n=1$, respectively. The influence of these modes on the transport dynamics can be quantified using the mean residence time $\tau_n$~\cite{cao_optimization_2009}. When computed in the eigenbasis of $H_\mathrm{ch}$, the mean residence time denotes the average time an excitation spends in a particular chain mode until decaying to the vacuum or trap state:
\begin{equation}
    \tau_n(t) = \int_0^t dt' \Tr{ \left\{ \rho(t') \ket{v_n}\bra{v_n} \right\} }.
\end{equation}
Fig.~\ref{fig:dephasing}(b) shows the mean residence time for each eigenmode as a function of the dephasing rate. In the case of $\ket{\Psi(0)} = \ket{G}$, the initially delocalized excitation couples strongly to the short-lived bright state. The corresponding contribution from dark state transport is negligible, and the excitation is rapidly lost to vacuum. More time spent in the bright state therefore leads to low transport efficiencies at zero dephasing. However, the difference in the mean residence times between the bright and dark states is reduced in the presence of dynamical disorder. In the site basis, the components of $\ket{B}$ and $\ket{D}$ are nearly identical in magnitude but differ in their phase relationships [Fig.~\ref{fig:dephasing}(c)]. As the strength of the thermal fluctuations is increased, the Lindbladian~\eqref{eq:dephasing} acts to randomize the phases of $\rho$ in the site basis with characteristic time $1/\Gamma_\phi$. The corresponding loss of coherence renders the bright and dark states largely indistinguishable. The excitation is then able to access the subradiant mode where it can reach the acceptor site before decaying. The loss of quantum coherence therefore enhances excitation transport by disrupting superradiant decay paths. 

On the other hand, when the initial state is fully localized in $\ket{e_1}$, Hamiltonian evolution generates large overlaps between the initial excitation and the long-lived dark state. The dark state is able to facilitate transport to the acceptor site with minimal losses, where it is subsequently trapped. When the dephasing rate is small, the time spent in the bright state is also small, and the excitation is protected from decaying to vacuum. As the dephasing strength is increased, the initial state generates overlap with the bright state and the transport efficiency is reduced. 

The inclusion of cooperative decay is quantitatively significant at small and intermediate dephasing ($\Gamma_\phi / J \lesssim 4$) where the mean residence times are different between the subradiant and superradiant collective modes. Transport for the localized initial state is less efficient in the independent decay model because the enhancement due to long dark state lifetimes is eliminated. The opposite effect is seen in the delocalized case due to the omission of the superradiant decay path. At larger dephasing, the excitation spends roughly equal amounts of time in states $\ket{B}$ and $\ket{D}$ such that the influence of cooperative decay is lost. This is an intuitive result: collective modes require coherent phase relationships that are randomized in the presence of strong fluctuations. Nevertheless, the off-diagonal elements of Eq.~\eqref{eq:G} cannot be neglected for dephasing rates on the order of the nearest-neighbor coupling strength (or smaller), especially when the initial state is delocalized (see also Appendix~\ref{app:collective}).

The above result is of significant importance to the study of biological photosynthetic energy transfer~\cite{mirkovic_light_2017, mattioni_design_2021}. In natural light-harvesting complexes, the inter-chromophore spacings are typically on the order of nanometers, whereas the optical transition wavelengths are in the visible range~\cite{jang_delocalized_2018}. These complexes are therefore extremely subwavelength and experience a symmetric phase distribution in response to incident light. In the absence of vibrational dephasing, the initial excitation would be in the form of the symmetric bright state that radiates rapidly to vacuum. The near perfect trapping efficiency observed in biological photosynthesis would therefore not be possible for such subwavelength complexes without strong vibrational coupling. Thus, the quantum-to-classical transition is explicitly manifest in biological energy transfer through the damping of bright state coherences.

\begin{figure*}
\centering  
\renewcommand\figurename{FIG.}
\includegraphics[width=\textwidth]{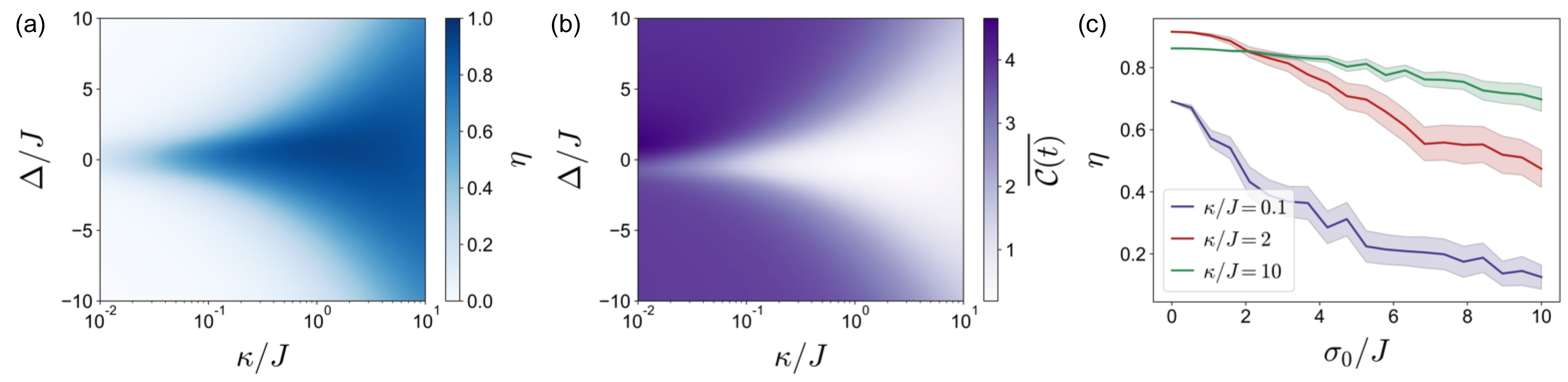}
\caption{(a) Trapping efficiency as a function of trapping rate $\kappa$ and acceptor detuning $\Delta$. Larger trapping rates result in a broader bandwidth. (b) Time-averaged coherent $\mathcal{C}(t)$ as a function of trapping rate and acceptor detuning. Optimal transport is associated with minimal coherence. (c) Average trapping efficiency across 100 realizations of $\Delta \in \mathcal{N}(0, \sigma_0)$. Larger trapping rates result in more robustness to disorder. Shaded uncertainty bands show 30\% distribution around the mean. Parameters for all panels: $N=10, a = 0.05 \lambda_0, t = 10/ \gamma$.}
\label{fig:broad}
\end{figure*}

Interestingly, we report a slight dephasing assisted transport enhancement for the delocalized state even in the absence of collective dissipation. This effect cannot be caused by the disruption of superradiant pathways because all eigenmodes have equal decay rates when $\Gamma_{ij} = \gamma \delta_{ij}$. We attribute this enhancement to classical homogeneous broadening induced by the dephasing process~\cite{caruso_highly_2009}. Because the initial state has only a very small overlap with the acceptor site, the excitation must make transitions to other states in order to reach the trap. However, the initial superposition state experiences an energy shift due to the coherent couplings $J_{ij}$ in the Hamiltonian. This pushes the initial state off resonance with the other available modes. In the presence of thermal fluctuations, the linewidths of these states are increased, with the homogeneous linewidth determined by $\Gamma_\phi$. Fig.~\ref{fig:dephasing}(d) shows the optimal dephasing rate for the delocalized initial state in the independent decay model under various lattice spacings. Values are plotted as a function of the average chain eigenmode energy splitting
\begin{equation}
    \overline{\Delta \varepsilon} = \sum_{n < m} \abs{\varepsilon_n - \varepsilon_m}
\end{equation}
for each geometry. The optimal dephasing rate increases linearly with the average energy splitting, suggesting an optimal spectral overlap resulting from homogeneous broadening. Importantly, however, this mechanism is not sufficient to explain the magnitude of the transport enhancement in the cooperative decay model. 

\subsection{The influence of static disorder\label{sect:static}}

As a final result, we now consider the influence of static disorder in the resonance frequency of the acceptor emitter. As a consequence of the additional lifetime broadening induced by the trapping process, highly efficient energy extraction can be achieved in both the broad-band and narrow-band regimes: a desirable property for quantum device design. The bandwidth of this process is determined by the trapping frequency and can be assessed by adding a frequency detuning to the acceptor emitter
\begin{equation}
    H_\Delta = \Delta \sigma_N^\dag \sigma_N.
\end{equation}
Fig.~\ref{fig:broad}(a) shows the trapping efficiency for the $N=10$ chain beginning with a localized excitation at the first site. As the trapping frequency increases, the range of acceptor detunings for which there is highly efficient trapping also increases. This allows for flexibility in the energy extraction process, as excitations of multiple frequencies can be trapped without sacrificing efficiency. The high trapping efficiencies are, once again, mediated by a reduction in quantum coherence [and therefore also by a reduction in entanglement via Eq.~\eqref{eq:entanglement}]. Fig.~\ref{fig:broad}(b) shows the time-averaged coherence for the same trapping frequency and detuning ranges as in panel (a). Much like in Section~\ref{sect:long-range}, the increase in total coherence for detunings outside the optimal bandwidth results from wave-like oscillations following reflection at the acceptor site. In this case, reflection is caused not by suboptimal trapping rates, but by the level shift of the acceptor. At larger trapping rates, the increase in acceptor linewidth compensates for this shift, allowing the excitation to be trapped without reflection. 

In turn, the broader bandwidth induced by larger trapping rates leads to a robustness against static disorder. Fig.~\ref{fig:broad}(c) shows the average trapping efficiency across 100 realizations with $\Delta$ drawn from a normal distribution $\mathcal{N}(0, \sigma_0)$ of zero mean and standard deviation $\sigma_0$. At large disorder ($\sigma_0 \gtrsim 2 J$), the optimal trapping rate is no longer equal to the group velocity of the resonant chain eigenmode and is instead dictated by the bandwidth of the acceptor site. The broad-band nature of this site allows for robust transport and trapping even with disorder much larger than the coherent coupling strength. This feature may allow for flexibility in designing incoherent energy extraction mechanisms that may introduce large amounts of local disorder.

\section{Conclusion\label{sect:conclusion}}

In this work, we have shown that a wide variety of decoherence mechanisms can be leveraged to enhance the extraction of excitonic energy from a paradigmatic quantum optical system. In particular, we have demonstrated that excitation trapping is optimized under conditions that minimize the total quantum coherence and entanglement of the system. This is in stark contrast to quantum random walk models of excitation transport that have demonstrated exponential speedups in information transfer. The key difference between those models and the results presented here is the explicit inclusion of the trapping process that mediates the retrieval of energy from the system. As shown, this trapping facilitates a quantum-to-classical transition through spontaneous $\PT$ symmetry breaking that results in unidirectional energy flow. We have also demonstrated that vibrational fluctuations can greatly enhance the trapping of delocalized excitations in the presence of cooperative radiative decay. This cooperative dissipation is present in all dipole coupled systems and cannot be neglected unless the dephasing strength is sufficiently large. Finally, we have examined the influence of static frequency disorder and shown that the transition from narrow-band to broad-band trapping results in an increased robustness to disorder, and is again associated with minimal quantum coherence. Due to the simplicity of the models studied, these phenomena are expected to be generalizable to arbitrary configurations of quantum emitters in one, two, and three dimensions.

Our results are of fundamental interest to the design of new photonics devices, artificial light-harvesting technologies, and to the study of biological photosynthetic energy transfer. Consideration of cooperative decay is important for the accurate quantification of transport efficiency, coherence, and entanglement in these systems, especially at subwavelength distances. For biological light-harvesting in particular, the high density packing of chromophore molecules into an extremely subwavelength volume presents a challenge for optimizing the transport efficiency due to the formation of short-lived bright states. On the one hand, a densely packed ensemble increases the chances of absorbing an incident photon. However, if the chromophores are too close together, they will be excited into the symmetric bright state and radiate strongly to vacuum before the absorbed photon can reach the reaction center. Our results suggest that vibrational dephasing is essential for disrupting bright state emission and can allow for increased transport efficiency while maintaining subwavelength scales. The ability to simultaneously maximize the number of photon absorbers while maintaining efficient excitation transport is a remarkable consequence of the quantum-to-classical transition applied to biological photosynthesis. Moreover, despite numerous studies examining photosynthetic energy transfer, comparatively little attention has been given to the role of excitation trapping at the photochemical reaction center. In addition to vibrational dephasing and static frequency disorder, the trapping rate represents another potential evolutionary knob that may have been tuned towards optimal energy extraction. Future studies of more realistic biological and biologically-inspired geometries can help assess the role of these interactions in nature and will help shed light on the many fascinating features associated with decoherence and the quantum-to-classical transition.

\section*{Acknowledgements}

R.H. acknowledges funding in whole or in part by the Austrian Science Fund (FWF) 10.55776/W1259. S.O. is supported by a postdoctoral fellowship of the Max Planck-Harvard Research Center for Quantum Optics. J.S.P., S.O., and S.F.Y. acknowledge funding from the National Science Foundation (NSF) via the Center for Ultracold Atoms (CUA) Physics Frontiers Centers (PFC) program and via PHY-2207972, as well as from the Air Force Office of Scientific Research (AFOSR).

\onecolumngrid
\appendix

\section{Light-matter interactions\label{app:light-matter}}

For a collection of $N$ two-level atoms with excited states $\ket{e_i}$, ground states $\ket{g_i}$, and resonance frequency $\omega_0$, the full atom-photon Hamiltonian is given in the electric dipole approximation by
\begin{equation}
    H_{\mathrm{at-ph}} = \sum_{i=1}^N \omega_0 \sigma^\dag_i \sigma_i + \sum_\k \nu_\k \left (a_\k^\dag a_\k + \frac{1}{2} \right ) - \sum_{i=1}^N \p_i \cdot \E.
\end{equation}
Here, $\sigma_i^\dag = \ket{e_i}\bra{g_i}$ and $\sigma_i = \ket{g_i}\bra{e_i}$ are atomic raising and lowering operators, and $a_\k^\dag$ ($a_\k$) creates (annihilates) a photon in mode $\k$ with frequency $\nu_\k$. The matter operators obey spin-1/2 commutation relations, though in the single excitation limit are equivalent to those of bosons or spinless fermions. The third term describes electric dipole interactions between the quantized electric field $\E = \sum_\k (\pmb{\mathcal{E}}_\k a_\k + \pmb{\mathcal{E}}^*_\k a_\k^\dag)$ and atoms with dipole moment $\p_i = \bwp_i \sigma_i + \bwp^*_i \sigma_i^\dag$. It is convenient to trace out the field degrees of freedom and to consider effective dipole-dipole interactions between the atoms directly. Treating the radiation field as a Markovian bath and applying the Born and rotating wave approximations in the usual way~\cite{lehmberg_radiation_1970, lehmberg_radiation_1970-1, asenjo-garcia_exponential_2017, breuer_theory_2010, scully_quantum_2008}, the system Hamiltonian and radiation Lindbladian can be written as
\begin{align}
    H &= \sum_{i=1}^N \omega_{0} \sigma_i^\dag \sigma_i + \sum_{i\neq j} J_{ij} \sigma_i^\dag \sigma_j \\
    \mathcal{L}_R[\rho] &= \sum^N_{i,j=1} \frac{\Gamma_{ij}}{2} [1 + \bar{n}(\omega_0)] \left ( 2 \sigma_j \rho \sigma_{i}^\dag - \left \{ \sigma_{i}^\dag \sigma_j, \rho \right \} \right ) + \sum^N_{i,j=1} \frac{\Gamma_{ij}}{2} \bar{n}(\omega_0) \left ( 2 \sigma^\dag_j \rho \sigma_{i} - \left \{ \sigma_{i} \sigma^\dag_j, \rho \right \} \right ),
\end{align}
where $\bar{n}(\omega_0) = 1/[e^{(\omega_0/(k_B T))} - 1]$ is the average thermal photon occupation in the Bose-Einstein distribution, $T$ is the temperature, and $k_B$ is the Boltzmann constant. For optical frequencies at room temperature ($\omega_0 \gg k_B T$), we may approximate $\bar{n}(\omega_0) \approx 0$. The quantities $J_{ij}$ and $\Gamma_{ij}$ describe the coherent and dissipative components of the effective interaction between oscillating optically-induced dipoles. They are given in terms of the dipole matrix element vector $\bwp_i = \wp_i \phat$ and the relative coordinate $\xi = \omega_0 r/c$ for $\mathbf{r} = r \rhat$, and $r = \abs{\r_i - \r_j}$ as
\begin{align}
    J_{ij} &=  \frac{3 \gamma}{4} \left\{ [3(\phat_i \cdot \rhat)(\phat_j \cdot \rhat) - \phat_i \cdot \phat_j]  \left(\frac{\cos{\xi}}{\xi^3} + \frac{\sin{\xi}}{\xi^2} \right) \right.
    \left. - [(\phat_i \cdot \rhat)(\phat_j \cdot \rhat) - \phat_i \cdot \phat_j] \left(\frac{\cos{\xi}}{\xi} \right) \right \} \\
    \Gamma_{ij} &=  \frac{3 \gamma}{2} \left\{ [3(\phat_i \cdot \rhat)(\phat_j \cdot \rhat) - \phat_i \cdot \phat_j]  \left(\frac{\sin{\xi}}{\xi^3} - \frac{\cos{\xi}}{\xi^2} \right) \right.
    \left. - [(\phat_i \cdot \rhat)(\phat_j \cdot \rhat) - \phat_i \cdot \phat_j] \left(\frac{\sin{\xi}}{\xi} \right) \right \}. \label{eq:app_G}
\end{align}
The terms proportional to $1/\xi^3$, $1/\xi^2$, and $1/\xi$ in each expression correspond to the familiar near field zone, intermediate zone, and radiation zone regimes of the time dependent electric dipole field. The term $\gamma \equiv \Gamma_{ii} = \omega_0^3 \abs{\bwp_i}^2 / (3 \pi \hbar \epsilon_0 c^3)$ describes the Wigner-Weisskopf spontaneous emission rate of each atom.

For very short distances ($\xi \ll 1$), or equivalently very short times ($\gamma t \ll 1$), it is common to invoke the quasistatic approximation and neglect the intermediate and radiation zone contributions. The atomic interactions are then reduced to those between electrostatic dipoles,
\begin{align}
    J_{ij} & \to \frac{3(\bwp^*_i \cdot \rhat)(\bwp_j \cdot \rhat) - \bwp^*_i \cdot \bwp_j}{4 \pi \epsilon_0 r^3},\\
    \Gamma_{ij} & \to \gamma \phat_i \cdot \phat_j.
\end{align}
Numerous studies, including those concerning photosynthesis and artificial light-harvesting, further neglect off-diagonal dissipation altogether by setting $\Gamma_{ij} = \delta_{ij} \gamma$. This approximation is, however, not generally correct when the exciton states are delocalized, even in the limit as $\xi \to 0$~\cite{shatokhin_coherence_2018}. The finite contribution of the off-diagonal $\Gamma_{ij}$ terms leads to cooperative phenomena, such as superradiance and subradiance, that are not captured when these terms are neglected (see Appendix~\ref{app:collective} below). Such effects are crucial to the accurate description of quantum optical systems, but are not typically included in transport analyses of biologically-inspired light-harvesting. 

For a general multi-excitation system, the time dynamics are governed by the master equation $\dot{\rho} = -i [H, \rho] + \mathcal{L}_R[\rho]$. However, if we restrict ourselves to the case when there is at most a single excitation in the system at any given time, we may neglect the quantum jump terms $\propto 2 \sigma_j \rho \sigma_i^\dag$ in the radiation Lindbladian. Setting $\bar{n}(\omega_0) = 0$, the equations of motion are then equivalent to evolving with the Schrodinger equation for the non-Hermitian effective Hamiltonian
\begin{equation}\label{eq:effective}
    \Heff = \sum_{i=1}^N \left(\omega_0 - \frac{i}{2} \gamma \right) \sigma_i^\dag \sigma_i + \sum_{i \neq j} \left(J_{ij} - \frac{i}{2} \Gamma_{ij} \right) \sigma_i^\dag \sigma_j.
\end{equation}
This single-excitation regime is most suitable to the study of optimal transport parameters under low light conditions, or when the rate of photon absorption is small relative to the inverse transport time.

\section{The role of collective dissipation\label{app:collective}}

\begin{figure}
\centering 
\renewcommand\figurename{FIG.}
\includegraphics[width=\textwidth]{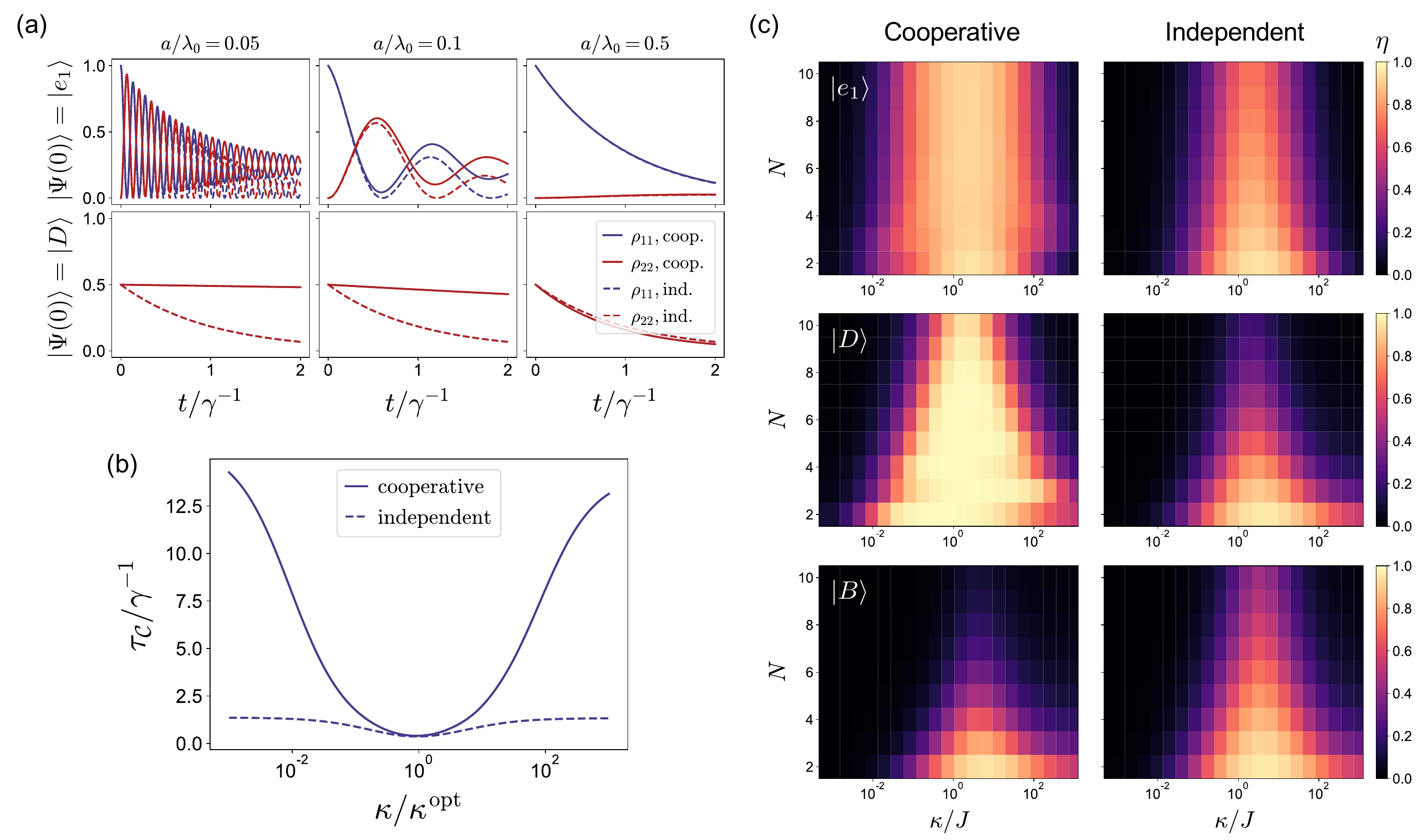}
\caption{(a) Site populations $\rho_{11}$ (blue) and $\rho_{22}$ (red) for the $N=2$ chain according to the effective Hamiltonian~\eqref{eq:effective}. Solid lines denote the full cooperative decay model [Eq.~\eqref{eq:app_G}] whereas dashed lines indicate the independent decay approximation $\Gamma_{ij} = \gamma \delta_{ij}$. Columns indicate different emitter spacings. The top row corresponds to an initial excitation on the first site, whereas the bottom row shows results when initializing in the dark state $\ket{D} = (\ket{e_1} - \ket{e_2}) / \sqrt{2}$. (b) The characteristic coherence time $\tau_\mathcal{C}$ for the $N=10$ chain of Section~\ref{sect:long-range} in the presence (solid line) and absence (dashed line) of cooperative decay. (c) Comparison of the trapping efficiency $\eta$ with (left) and without (right) cooperative decay as a function of the trapping rate and chain length. Different rows correspond to different initial states. From top to bottom: $\ket{\Psi(0) = \ket{e_1}}$, $\ket{\Psi(0) = \ket{D}}$, $\ket{\Psi(0) = \ket{B}}$. Additional parameters for panels (b) and (c): $a/\lambda_0 = 0.05, \gamma t = 10$.}
\label{fig:collective}
\end{figure}

Here we expand on the necessity of including off-diagonal collective dissipation in order to accurately describe excitation transport between quantum emitters. It is sometimes assumed that for inter-emitter spacing $a$ much less than the dipole transition wavelength $\lambda_0$ that the off-diagonal terms of the dissipative interaction $\Gamma_{ij}$ can be neglected. This erroneous assumption likely stems from the fact that $J_{ij} \gg \Gamma_{ij}$ when $a \ll \lambda_0$, but is incorrect because it neglects the role of destructive interference. Fig.~\ref{fig:collective}(a) compares the population dynamics for a two emitter system in the presence [solid lines, $\Gamma_{ij}$ determined by Eq.~\eqref{eq:app_G}] and absence (dashed lines, $\Gamma_{ij} = \delta_{ij}\gamma$) of cooperative decay. In direct contrast with the above assumption, the discrepancy between the cooperative and independent decay models actually increases as $a / \lambda_0 \to 0$. For subwavelength spacings, the approximation $\Gamma_{ij} = \gamma \delta_{ij}$ is valid only for localized excitations at very short times ($t \ll \gamma^{-1}$). For delocalized excitations, the approximation is applicable only when the emitters are far apart, or in the presence of strong dephasing (Fig.~\ref{fig:dephasing}).

The independent decay model can be used in the presence of strong dephasing because dephasing destroys the phase coherences required for destructive interference. However, as discussed in the main text, these coherences play an instrumental role in describing the fundamental aspects of excitation transport and trapping. Fig.~\ref{fig:collective}(b) compares the coherence time results of Section~\ref{sect:long-range} with the corresponding values in the independent decay model for the $N=10$ chain. The transition between the quantum and classical regimes is obfuscated by the neglect of cooperative dissipation. Moreover, this approximation also prohibits an accurate quantitative assessment of the trapping efficiency. Fig.~\ref{fig:collective}(c) shows the results with and without cooperative dissipation when initializing in either the dark or bright state. When collective effects are included, initializing in $\ket{D}$ leads to minimal losses and enhances the transport as compared to the localized initial state. The opposite effect is seen for the bright state $\ket{B}$. In the independent decay model, all modes have equal decay rates and the discrepancy between the dark and bright states is greatly reduced. Collective dissipation is therefore essential to the accurate description of excitation trapping in the absence of strong vibrational fluctuations.  

\section{$\PT$ symmetry breaking in the two-site model\label{app:PT}}

The $\PT$ symmetry breaking transition is most easily observed through the gauge transformation
\begin{equation}
    \rcvect{c_1\\c_2} \rightarrow e^{-\chi t/2} \rcvect{c_1\\c_2}
\end{equation}
which removes the trivial global excitation loss with rate $\chi$. The resulting gauge-transformed Hamiltonian $H'_{\mathrm{eff}}$ obeys $[\mathcal{PT}, H_{\mathrm{eff}}'] = 0$, where, for the bipartite system, the parity and time-reversal operators act as $\mathcal{P} = \sigma_x$ and $\mathcal{T}: i \to -i$. $H'$ has eigenvectors
\begin{equation}\label{eq:2_evects}
\ket{v_\pm} = \frac{1}{\sqrt{1 + \abs{\frac{i\kappa \pm 4\lambda}{4J}}^2}} \rcvect{
\frac{i\kappa \pm 4\lambda}{4J}\\
1
}
\end{equation}
and eigenvalues $\tilde{\varepsilon}_\pm = \pm \lambda$ that transition from purely real to purely imaginary when $\kappa = 4J$. For $4J > \kappa$, $\lambda \in \mathbb{R}$ such that $\abs{(i\kappa \pm 4\lambda) / 4J} = 1$ and Eq.~\eqref{eq:2_evects} may be written as
\begin{equation}
\ket{v_\pm} = \frac{e^{i\alpha_\pm/2}}{\sqrt{2}} \rcvect{
e^{i\alpha_\pm/2}\\e^{-i\alpha_\pm/2}}
\end{equation}
for $\alpha_\pm = \mathrm{arg}[(i\kappa \pm 4\lambda) / 4J]$. It is easy to verify in this case that $\mathcal{PT}: \ket{v_\pm} \rightarrow e^{-i\alpha_\pm} \ket{v_\pm}$ such that the eigenstates are invariant under the combined $\mathcal{PT}$ operation up to a $U(1)$ gauge ambiguity. For $4J < \kappa$, the components of $\ket{v_\pm}$ no longer have equal modulus and the $\mathcal{PT}$ symmetry is spontaneously broken.

\section{EPs of the nearest-neighbor Hamiltonian}\label{app:EP}

The nearest-neighbor effective Hamiltonian given in Eq.~\eqref{eq:nn} can be written as $\Heff = H_0 + H'$, where
\begin{align}
    H_0 &= \sum_{i=1}^N \left (\omega_0 - \frac{i}{2} \gamma \right) \sigma_i^\dag \sigma_i,\\
    H' &=J \sum_{\braket{i,j}} \sigma_i^\dag \sigma_j - \frac{i}{2} \kappa \sigma_N^\dag \sigma_N.
\end{align}
The EPs of the system can be found by first solving for the eigenvalues of $\Heff$ that have algebraic multiplicity greater than their geometric multiplicity. To aid in this calculation, we first move to the interaction picture with $H_0$ and define the interaction picture states $\ket{\hat{\Psi}(t)} = e^{iH_0t} \ket{\Psi(t)}$. In the single excitation subspace, the number operator $\sigma_i^\dag \sigma_i$ simply acts as the identity, and the interaction Hamiltonian satisfies $\hat{H}' = e^{iH_0t} H' e^{-iH_0t} = H'$. The eigenstates then obey $\ket{\hat{v}_n} = e^{(i\omega_0 + \gamma/2)t} \ket{v_n}$. In other words, the eigenstates are transformed into the frame rotating with complex frequency $\omega_0 - i\gamma/2$. This change of reference frame does not change the multiplicity of the eigenvalues, meaning we can work exclusively with $H'$. For $\kappa=0$, the eigenvalues of $H'$ are distinct and given by $\tilde{\varepsilon}_n = 2J\cos{[n\pi/(N+1)]}$ with orthonormal eigenvectors $\ket{v_n} = \sum_j v_{nj} \ket{e_j}$, where $v_{nj} = \sin{[nj\pi/(N+1)]}$ and $n,j = 1,...,N$. This is the standard result for a finite-size tight-binding Hamiltonian \cite{noschese_tridiagonal_2013}. However for $\kappa \neq 0$, the eigenvalues of $H'$ are \emph{not} necessarily distinct, but can still be calculated as the roots of the characteristic polynomial $\phi_N(\tilde{\varepsilon}) = \det(H' - \tilde{\varepsilon} \openone_N)$, where $\openone_N$ is the $N \times N$ identity matrix. In search of an analytic solution, we note that the Hamiltonian may be written as $H'/J = T - (i\kappa/2J) K$, where $T_{ij} = \delta_{i,j+1} + \delta_{i+1,j}$ is an $N \times N$ tridiagonal symmetric Toeplitz matrix with zeros on the diagonal and $K_{ij} = \delta_{iN}\delta_{Nj}$. Using the identities derived in Ref.~\cite{kulkarni_eigenvalues_1999}, the characteristic polynomial for such a matrix is given by
\begin{equation}
    \phi_N(\tilde{\varepsilon}) = \left(\frac{-i\kappa}{2J} - \frac{\tilde{\varepsilon}}{J} \right) U_{N-1} \left(\frac{-\tilde{\varepsilon}}{2} \right) - U_{N-2} \left(\frac{-\tilde{\varepsilon}}{2} \right)
\end{equation}
where $U_n(x)$ is the $n$\ts{th} degree Chebyshev polynomial of the second kind. Finally, using the identity $\det(\alpha A - \tilde{\varepsilon} \openone_N) = \alpha^N \det(A - (\tilde{\varepsilon}/\alpha) \openone_N)$ for scalar $\alpha$ and $N \times N$ matrix $A$, the characteristic polynomial for $H'$ is given by
\begin{equation}
    \phi_N(\tilde{\varepsilon}) = J^N \left[ \left(\frac{-i\kappa}{2J} - \frac{\tilde{\varepsilon}}{J} \right) U_{N-1} \left (\frac{-\tilde{\varepsilon}}{2J} \right ) - U_{N-2} \left (\frac{-\tilde{\varepsilon}}{2J} \right) \right],
\end{equation}
in agreement with Eq.~\eqref{eq:char_poly}.

\bibliography{refs.bib}

\begin{thebibliography}{63}%
\makeatletter
\providecommand \@ifxundefined [1]{%
 \@ifx{#1\undefined}
}%
\providecommand \@ifnum [1]{%
 \ifnum #1\expandafter \@firstoftwo
 \else \expandafter \@secondoftwo
 \fi
}%
\providecommand \@ifx [1]{%
 \ifx #1\expandafter \@firstoftwo
 \else \expandafter \@secondoftwo
 \fi
}%
\providecommand \natexlab [1]{#1}%
\providecommand \enquote  [1]{``#1''}%
\providecommand \bibnamefont  [1]{#1}%
\providecommand \bibfnamefont [1]{#1}%
\providecommand \citenamefont [1]{#1}%
\providecommand \href@noop [0]{\@secondoftwo}%
\providecommand \href [0]{\begingroup \@sanitize@url \@href}%
\providecommand \@href[1]{\@@startlink{#1}\@@href}%
\providecommand \@@href[1]{\endgroup#1\@@endlink}%
\providecommand \@sanitize@url [0]{\catcode `\\12\catcode `\$12\catcode
  `\&12\catcode `\#12\catcode `\^12\catcode `\_12\catcode `\%12\relax}%
\providecommand \@@startlink[1]{}%
\providecommand \@@endlink[0]{}%
\providecommand \url  [0]{\begingroup\@sanitize@url \@url }%
\providecommand \@url [1]{\endgroup\@href {#1}{\urlprefix }}%
\providecommand \urlprefix  [0]{URL }%
\providecommand \Eprint [0]{\href }%
\providecommand \doibase [0]{https://doi.org/}%
\providecommand \selectlanguage [0]{\@gobble}%
\providecommand \bibinfo  [0]{\@secondoftwo}%
\providecommand \bibfield  [0]{\@secondoftwo}%
\providecommand \translation [1]{[#1]}%
\providecommand \BibitemOpen [0]{}%
\providecommand \bibitemStop [0]{}%
\providecommand \bibitemNoStop [0]{.\EOS\space}%
\providecommand \EOS [0]{\spacefactor3000\relax}%
\providecommand \BibitemShut  [1]{\csname bibitem#1\endcsname}%
\let\auto@bib@innerbib\@empty
\bibitem [{\citenamefont {Childs}\ \emph {et~al.}(2002)\citenamefont {Childs},
  \citenamefont {Farhi},\ and\ \citenamefont {Gutmann}}]{childs_example_2002}%
  \BibitemOpen
  \bibfield  {author} {\bibinfo {author} {\bibfnamefont {A.~M.}\ \bibnamefont
  {Childs}}, \bibinfo {author} {\bibfnamefont {E.}~\bibnamefont {Farhi}},\ and\
  \bibinfo {author} {\bibfnamefont {S.}~\bibnamefont {Gutmann}},\ }\bibfield
  {title} {\bibinfo {title} {An example of the difference between quantum and
  classical random walks},\ }\href {https://doi.org/10.1023/A:1019609420309}
  {\bibfield  {journal} {\bibinfo  {journal} {Quantum Information Processing}\
  }\textbf {\bibinfo {volume} {1}},\ \bibinfo {pages} {35} (\bibinfo {year}
  {2002})}\BibitemShut {NoStop}%
\bibitem [{\citenamefont {Higgins}\ \emph {et~al.}(2014)\citenamefont
  {Higgins}, \citenamefont {Benjamin}, \citenamefont {Stace}, \citenamefont
  {Milburn}, \citenamefont {Lovett},\ and\ \citenamefont
  {Gauger}}]{higgins_superabsorption_2014}%
  \BibitemOpen
  \bibfield  {author} {\bibinfo {author} {\bibfnamefont {K.~D.~B.}\
  \bibnamefont {Higgins}}, \bibinfo {author} {\bibfnamefont {S.~C.}\
  \bibnamefont {Benjamin}}, \bibinfo {author} {\bibfnamefont {T.~M.}\
  \bibnamefont {Stace}}, \bibinfo {author} {\bibfnamefont {G.~J.}\ \bibnamefont
  {Milburn}}, \bibinfo {author} {\bibfnamefont {B.~W.}\ \bibnamefont
  {Lovett}},\ and\ \bibinfo {author} {\bibfnamefont {E.~M.}\ \bibnamefont
  {Gauger}},\ }\bibfield  {title} {\bibinfo {title} {Superabsorption of light
  via quantum engineering},\ }\bibfield  {journal} {\bibinfo  {journal} {Nature
  Communications}\ }\textbf {\bibinfo {volume} {5}},\ \href
  {https://doi.org/10.1038/ncomms5705} {10.1038/ncomms5705} (\bibinfo {year}
  {2014})\BibitemShut {NoStop}%
\bibitem [{\citenamefont {Nelson}(2018)}]{nelson_role_2018}%
  \BibitemOpen
  \bibfield  {author} {\bibinfo {author} {\bibfnamefont {P.~C.}\ \bibnamefont
  {Nelson}},\ }\bibfield  {title} {\bibinfo {title} {The {Role} of {Quantum}
  {Decoherence} in {FRET}},\ }\href {https://doi.org/10.1016/j.bpj.2018.01.010}
  {\bibfield  {journal} {\bibinfo  {journal} {Biophysical Journal}\ }\textbf
  {\bibinfo {volume} {115}},\ \bibinfo {pages} {167} (\bibinfo {year}
  {2018})}\BibitemShut {NoStop}%
\bibitem [{\citenamefont {Schlosshauer}(2005)}]{schlosshauer_decoherence_2005}%
  \BibitemOpen
  \bibfield  {author} {\bibinfo {author} {\bibfnamefont {M.}~\bibnamefont
  {Schlosshauer}},\ }\bibfield  {title} {\bibinfo {title} {Decoherence, the
  measurement problem, and interpretations of quantum mechanics},\ }\href
  {https://doi.org/10.1103/RevModPhys.76.1267} {\bibfield  {journal} {\bibinfo
  {journal} {Reviews of Modern Physics}\ }\textbf {\bibinfo {volume} {76}},\
  \bibinfo {pages} {1267} (\bibinfo {year} {2005})}\BibitemShut {NoStop}%
\bibitem [{\citenamefont {Schlosshauer}(2007)}]{schlosshauer_decoherence_2007}%
  \BibitemOpen
  \bibfield  {author} {\bibinfo {author} {\bibfnamefont {M.~A.}\ \bibnamefont
  {Schlosshauer}},\ }\href@noop {} {\emph {\bibinfo {title} {Decoherence and
  the quantum-to-classical transition}}},\ The frontiers collection\ (\bibinfo
  {publisher} {Springer},\ \bibinfo {address} {Berlin ; London},\ \bibinfo
  {year} {2007})\BibitemShut {NoStop}%
\bibitem [{\citenamefont {Zurek}(2003)}]{zurek_decoherence_2003}%
  \BibitemOpen
  \bibfield  {author} {\bibinfo {author} {\bibfnamefont {W.~H.}\ \bibnamefont
  {Zurek}},\ }\bibfield  {title} {\bibinfo {title} {Decoherence, einselection,
  and the quantum origins of the classical},\ }\href
  {https://doi.org/10.1103/RevModPhys.75.715} {\bibfield  {journal} {\bibinfo
  {journal} {Reviews of Modern Physics}\ }\textbf {\bibinfo {volume} {75}},\
  \bibinfo {pages} {715} (\bibinfo {year} {2003})}\BibitemShut {NoStop}%
\bibitem [{\citenamefont {Anderson}(1958)}]{anderson_absence_1958}%
  \BibitemOpen
  \bibfield  {author} {\bibinfo {author} {\bibfnamefont {P.~W.}\ \bibnamefont
  {Anderson}},\ }\bibfield  {title} {\bibinfo {title} {Absence of {Diffusion}
  in {Certain} {Random} {Lattices}},\ }\href
  {https://doi.org/10.1103/PhysRev.109.1492} {\bibfield  {journal} {\bibinfo
  {journal} {Physical Review}\ }\textbf {\bibinfo {volume} {109}},\ \bibinfo
  {pages} {1492} (\bibinfo {year} {1958})}\BibitemShut {NoStop}%
\bibitem [{\citenamefont {Maier}\ \emph {et~al.}(2019)\citenamefont {Maier},
  \citenamefont {Brydges}, \citenamefont {Jurcevic}, \citenamefont {Trautmann},
  \citenamefont {Hempel}, \citenamefont {Lanyon}, \citenamefont {Hauke},
  \citenamefont {Blatt},\ and\ \citenamefont
  {Roos}}]{maier_environment-assisted_2019}%
  \BibitemOpen
  \bibfield  {author} {\bibinfo {author} {\bibfnamefont {C.}~\bibnamefont
  {Maier}}, \bibinfo {author} {\bibfnamefont {T.}~\bibnamefont {Brydges}},
  \bibinfo {author} {\bibfnamefont {P.}~\bibnamefont {Jurcevic}}, \bibinfo
  {author} {\bibfnamefont {N.}~\bibnamefont {Trautmann}}, \bibinfo {author}
  {\bibfnamefont {C.}~\bibnamefont {Hempel}}, \bibinfo {author} {\bibfnamefont
  {B.~P.}\ \bibnamefont {Lanyon}}, \bibinfo {author} {\bibfnamefont
  {P.}~\bibnamefont {Hauke}}, \bibinfo {author} {\bibfnamefont
  {R.}~\bibnamefont {Blatt}},\ and\ \bibinfo {author} {\bibfnamefont {C.~F.}\
  \bibnamefont {Roos}},\ }\bibfield  {title} {\bibinfo {title}
  {Environment-{Assisted} {Quantum} {Transport} in a 10-qubit {Network}},\
  }\href {https://doi.org/10.1103/PhysRevLett.122.050501} {\bibfield  {journal}
  {\bibinfo  {journal} {Physical Review Letters}\ }\textbf {\bibinfo {volume}
  {122}},\ \bibinfo {pages} {050501} (\bibinfo {year} {2019})}\BibitemShut
  {NoStop}%
\bibitem [{\citenamefont {Rebentrost}\ \emph
  {et~al.}(2009{\natexlab{a}})\citenamefont {Rebentrost}, \citenamefont
  {Mohseni}, \citenamefont {Kassal}, \citenamefont {Lloyd},\ and\ \citenamefont
  {Aspuru-Guzik}}]{rebentrost_environment-assisted_2009}%
  \BibitemOpen
  \bibfield  {author} {\bibinfo {author} {\bibfnamefont {P.}~\bibnamefont
  {Rebentrost}}, \bibinfo {author} {\bibfnamefont {M.}~\bibnamefont {Mohseni}},
  \bibinfo {author} {\bibfnamefont {I.}~\bibnamefont {Kassal}}, \bibinfo
  {author} {\bibfnamefont {S.}~\bibnamefont {Lloyd}},\ and\ \bibinfo {author}
  {\bibfnamefont {A.}~\bibnamefont {Aspuru-Guzik}},\ }\bibfield  {title}
  {\bibinfo {title} {Environment-assisted quantum transport},\ }\href
  {https://doi.org/10.1088/1367-2630/11/3/033003} {\bibfield  {journal}
  {\bibinfo  {journal} {New Journal of Physics}\ }\textbf {\bibinfo {volume}
  {11}},\ \bibinfo {pages} {033003} (\bibinfo {year}
  {2009}{\natexlab{a}})}\BibitemShut {NoStop}%
\bibitem [{\citenamefont {Chin}\ \emph {et~al.}(2010)\citenamefont {Chin},
  \citenamefont {Datta}, \citenamefont {Caruso}, \citenamefont {Huelga},\ and\
  \citenamefont {Plenio}}]{chin_noise-assisted_2010}%
  \BibitemOpen
  \bibfield  {author} {\bibinfo {author} {\bibfnamefont {A.~W.}\ \bibnamefont
  {Chin}}, \bibinfo {author} {\bibfnamefont {A.}~\bibnamefont {Datta}},
  \bibinfo {author} {\bibfnamefont {F.}~\bibnamefont {Caruso}}, \bibinfo
  {author} {\bibfnamefont {S.~F.}\ \bibnamefont {Huelga}},\ and\ \bibinfo
  {author} {\bibfnamefont {M.~B.}\ \bibnamefont {Plenio}},\ }\bibfield  {title}
  {\bibinfo {title} {Noise-assisted energy transfer in quantum networks and
  light-harvesting complexes},\ }\href
  {https://doi.org/10.1088/1367-2630/12/6/065002} {\bibfield  {journal}
  {\bibinfo  {journal} {New Journal of Physics}\ }\textbf {\bibinfo {volume}
  {12}},\ \bibinfo {pages} {065002} (\bibinfo {year} {2010})}\BibitemShut
  {NoStop}%
\bibitem [{\citenamefont {Plenio}\ and\ \citenamefont
  {Huelga}(2008)}]{plenio_dephasing-assisted_2008}%
  \BibitemOpen
  \bibfield  {author} {\bibinfo {author} {\bibfnamefont {M.~B.}\ \bibnamefont
  {Plenio}}\ and\ \bibinfo {author} {\bibfnamefont {S.~F.}\ \bibnamefont
  {Huelga}},\ }\bibfield  {title} {\bibinfo {title} {Dephasing-assisted
  transport: quantum networks and biomolecules},\ }\href
  {https://doi.org/10.1088/1367-2630/10/11/113019} {\bibfield  {journal}
  {\bibinfo  {journal} {New Journal of Physics}\ }\textbf {\bibinfo {volume}
  {10}},\ \bibinfo {pages} {113019} (\bibinfo {year} {2008})}\BibitemShut
  {NoStop}%
\bibitem [{\citenamefont {Mohseni}\ \emph {et~al.}(2008)\citenamefont
  {Mohseni}, \citenamefont {Rebentrost}, \citenamefont {Lloyd},\ and\
  \citenamefont {Aspuru-Guzik}}]{mohseni_environment-assisted_2008}%
  \BibitemOpen
  \bibfield  {author} {\bibinfo {author} {\bibfnamefont {M.}~\bibnamefont
  {Mohseni}}, \bibinfo {author} {\bibfnamefont {P.}~\bibnamefont {Rebentrost}},
  \bibinfo {author} {\bibfnamefont {S.}~\bibnamefont {Lloyd}},\ and\ \bibinfo
  {author} {\bibfnamefont {A.}~\bibnamefont {Aspuru-Guzik}},\ }\bibfield
  {title} {\bibinfo {title} {Environment-assisted quantum walks in
  photosynthetic energy transfer},\ }\href {https://doi.org/10.1063/1.3002335}
  {\bibfield  {journal} {\bibinfo  {journal} {The Journal of Chemical Physics}\
  }\textbf {\bibinfo {volume} {129}},\ \bibinfo {pages} {174106} (\bibinfo
  {year} {2008})}\BibitemShut {NoStop}%
\bibitem [{\citenamefont {Rebentrost}\ \emph
  {et~al.}(2009{\natexlab{b}})\citenamefont {Rebentrost}, \citenamefont
  {Mohseni},\ and\ \citenamefont {Aspuru-Guzik}}]{rebentrost_role_2009}%
  \BibitemOpen
  \bibfield  {author} {\bibinfo {author} {\bibfnamefont {P.}~\bibnamefont
  {Rebentrost}}, \bibinfo {author} {\bibfnamefont {M.}~\bibnamefont
  {Mohseni}},\ and\ \bibinfo {author} {\bibfnamefont {A.}~\bibnamefont
  {Aspuru-Guzik}},\ }\bibfield  {title} {\bibinfo {title} {Role of {Quantum}
  {Coherence} and {Environmental} {Fluctuations} in {Chromophoric} {Energy}
  {Transport}},\ }\href {https://doi.org/10.1021/jp901724d} {\bibfield
  {journal} {\bibinfo  {journal} {The Journal of Physical Chemistry B}\
  }\textbf {\bibinfo {volume} {113}},\ \bibinfo {pages} {9942} (\bibinfo {year}
  {2009}{\natexlab{b}})}\BibitemShut {NoStop}%
\bibitem [{\citenamefont {Contreras-Pulido}\ \emph {et~al.}(2014)\citenamefont
  {Contreras-Pulido}, \citenamefont {Bruderer}, \citenamefont {Huelga},\ and\
  \citenamefont {Plenio}}]{contreras-pulido_dephasing-assisted_2014}%
  \BibitemOpen
  \bibfield  {author} {\bibinfo {author} {\bibfnamefont {L.~D.}\ \bibnamefont
  {Contreras-Pulido}}, \bibinfo {author} {\bibfnamefont {M.}~\bibnamefont
  {Bruderer}}, \bibinfo {author} {\bibfnamefont {S.~F.}\ \bibnamefont
  {Huelga}},\ and\ \bibinfo {author} {\bibfnamefont {M.~B.}\ \bibnamefont
  {Plenio}},\ }\bibfield  {title} {\bibinfo {title} {Dephasing-assisted
  transport in linear triple quantum dots},\ }\href
  {https://doi.org/10.1088/1367-2630/16/11/113061} {\bibfield  {journal}
  {\bibinfo  {journal} {New Journal of Physics}\ }\textbf {\bibinfo {volume}
  {16}},\ \bibinfo {pages} {113061} (\bibinfo {year} {2014})}\BibitemShut
  {NoStop}%
\bibitem [{\citenamefont {Contreras-Pulido}\ and\ \citenamefont
  {Bruderer}(2017)}]{contreras-pulido_coherent_2017}%
  \BibitemOpen
  \bibfield  {author} {\bibinfo {author} {\bibfnamefont {L.~D.}\ \bibnamefont
  {Contreras-Pulido}}\ and\ \bibinfo {author} {\bibfnamefont {M.}~\bibnamefont
  {Bruderer}},\ }\bibfield  {title} {\bibinfo {title} {Coherent and incoherent
  charge transport in linear triple quantum dots},\ }\href
  {https://doi.org/10.1088/1361-648X/aa66d0} {\bibfield  {journal} {\bibinfo
  {journal} {Journal of Physics: Condensed Matter}\ }\textbf {\bibinfo {volume}
  {29}},\ \bibinfo {pages} {185301} (\bibinfo {year} {2017})}\BibitemShut
  {NoStop}%
\bibitem [{\citenamefont {Engel}\ \emph {et~al.}(2007)\citenamefont {Engel},
  \citenamefont {Calhoun}, \citenamefont {Read}, \citenamefont {Ahn},
  \citenamefont {Mančal}, \citenamefont {Cheng}, \citenamefont {Blankenship},\
  and\ \citenamefont {Fleming}}]{engel_evidence_2007}%
  \BibitemOpen
  \bibfield  {author} {\bibinfo {author} {\bibfnamefont {G.~S.}\ \bibnamefont
  {Engel}}, \bibinfo {author} {\bibfnamefont {T.~R.}\ \bibnamefont {Calhoun}},
  \bibinfo {author} {\bibfnamefont {E.~L.}\ \bibnamefont {Read}}, \bibinfo
  {author} {\bibfnamefont {T.-K.}\ \bibnamefont {Ahn}}, \bibinfo {author}
  {\bibfnamefont {T.}~\bibnamefont {Mančal}}, \bibinfo {author} {\bibfnamefont
  {Y.-C.}\ \bibnamefont {Cheng}}, \bibinfo {author} {\bibfnamefont {R.~E.}\
  \bibnamefont {Blankenship}},\ and\ \bibinfo {author} {\bibfnamefont {G.~R.}\
  \bibnamefont {Fleming}},\ }\bibfield  {title} {\bibinfo {title} {Evidence for
  wavelike energy transfer through quantum coherence in photosynthetic
  systems},\ }\href {https://doi.org/10.1038/nature05678} {\bibfield  {journal}
  {\bibinfo  {journal} {Nature}\ }\textbf {\bibinfo {volume} {446}},\ \bibinfo
  {pages} {782} (\bibinfo {year} {2007})}\BibitemShut {NoStop}%
\bibitem [{\citenamefont {Ishizaki}\ \emph {et~al.}(2010)\citenamefont
  {Ishizaki}, \citenamefont {Calhoun}, \citenamefont {Schlau-Cohen},\ and\
  \citenamefont {Fleming}}]{ishizaki_quantum_2010}%
  \BibitemOpen
  \bibfield  {author} {\bibinfo {author} {\bibfnamefont {A.}~\bibnamefont
  {Ishizaki}}, \bibinfo {author} {\bibfnamefont {T.~R.}\ \bibnamefont
  {Calhoun}}, \bibinfo {author} {\bibfnamefont {G.~S.}\ \bibnamefont
  {Schlau-Cohen}},\ and\ \bibinfo {author} {\bibfnamefont {G.~R.}\ \bibnamefont
  {Fleming}},\ }\bibfield  {title} {\bibinfo {title} {Quantum coherence and its
  interplay with protein environments in photosynthetic electronic energy
  transfer},\ }\href {https://doi.org/10.1039/C003389H} {\bibfield  {journal}
  {\bibinfo  {journal} {Physical Chemistry Chemical Physics}\ }\textbf
  {\bibinfo {volume} {12}},\ \bibinfo {pages} {7319} (\bibinfo {year}
  {2010})}\BibitemShut {NoStop}%
\bibitem [{\citenamefont {Ishizaki}\ and\ \citenamefont
  {Fleming}(2012)}]{ishizaki_quantum_2012}%
  \BibitemOpen
  \bibfield  {author} {\bibinfo {author} {\bibfnamefont {A.}~\bibnamefont
  {Ishizaki}}\ and\ \bibinfo {author} {\bibfnamefont {G.~R.}\ \bibnamefont
  {Fleming}},\ }\bibfield  {title} {\bibinfo {title} {Quantum {Coherence} in
  {Photosynthetic} {Light} {Harvesting}},\ }\href
  {https://doi.org/10.1146/annurev-conmatphys-020911-125126} {\bibfield
  {journal} {\bibinfo  {journal} {Annual Review of Condensed Matter Physics}\
  }\textbf {\bibinfo {volume} {3}},\ \bibinfo {pages} {333} (\bibinfo {year}
  {2012})}\BibitemShut {NoStop}%
\bibitem [{\citenamefont {Scholes}\ \emph {et~al.}(2017)\citenamefont
  {Scholes}, \citenamefont {Fleming}, \citenamefont {Chen}, \citenamefont
  {Aspuru-Guzik}, \citenamefont {Buchleitner}, \citenamefont {Coker},
  \citenamefont {Engel}, \citenamefont {van Grondelle}, \citenamefont
  {Ishizaki}, \citenamefont {Jonas}, \citenamefont {Lundeen}, \citenamefont
  {McCusker}, \citenamefont {Mukamel}, \citenamefont {Ogilvie}, \citenamefont
  {Olaya-Castro}, \citenamefont {Ratner}, \citenamefont {Spano}, \citenamefont
  {Whaley},\ and\ \citenamefont {Zhu}}]{scholes_using_2017}%
  \BibitemOpen
  \bibfield  {author} {\bibinfo {author} {\bibfnamefont {G.~D.}\ \bibnamefont
  {Scholes}}, \bibinfo {author} {\bibfnamefont {G.~R.}\ \bibnamefont
  {Fleming}}, \bibinfo {author} {\bibfnamefont {L.~X.}\ \bibnamefont {Chen}},
  \bibinfo {author} {\bibfnamefont {A.}~\bibnamefont {Aspuru-Guzik}}, \bibinfo
  {author} {\bibfnamefont {A.}~\bibnamefont {Buchleitner}}, \bibinfo {author}
  {\bibfnamefont {D.~F.}\ \bibnamefont {Coker}}, \bibinfo {author}
  {\bibfnamefont {G.~S.}\ \bibnamefont {Engel}}, \bibinfo {author}
  {\bibfnamefont {R.}~\bibnamefont {van Grondelle}}, \bibinfo {author}
  {\bibfnamefont {A.}~\bibnamefont {Ishizaki}}, \bibinfo {author}
  {\bibfnamefont {D.~M.}\ \bibnamefont {Jonas}}, \bibinfo {author}
  {\bibfnamefont {J.~S.}\ \bibnamefont {Lundeen}}, \bibinfo {author}
  {\bibfnamefont {J.~K.}\ \bibnamefont {McCusker}}, \bibinfo {author}
  {\bibfnamefont {S.}~\bibnamefont {Mukamel}}, \bibinfo {author} {\bibfnamefont
  {J.~P.}\ \bibnamefont {Ogilvie}}, \bibinfo {author} {\bibfnamefont
  {A.}~\bibnamefont {Olaya-Castro}}, \bibinfo {author} {\bibfnamefont {M.~A.}\
  \bibnamefont {Ratner}}, \bibinfo {author} {\bibfnamefont {F.~C.}\
  \bibnamefont {Spano}}, \bibinfo {author} {\bibfnamefont {K.~B.}\ \bibnamefont
  {Whaley}},\ and\ \bibinfo {author} {\bibfnamefont {X.}~\bibnamefont {Zhu}},\
  }\bibfield  {title} {\bibinfo {title} {Using coherence to enhance function in
  chemical and biophysical systems},\ }\href
  {https://doi.org/10.1038/nature21425} {\bibfield  {journal} {\bibinfo
  {journal} {Nature}\ }\textbf {\bibinfo {volume} {543}},\ \bibinfo {pages}
  {647} (\bibinfo {year} {2017})}\BibitemShut {NoStop}%
\bibitem [{\citenamefont {Huelga}\ and\ \citenamefont
  {Plenio}(2013)}]{huelga_vibrations_2013}%
  \BibitemOpen
  \bibfield  {author} {\bibinfo {author} {\bibfnamefont {S.~F.}\ \bibnamefont
  {Huelga}}\ and\ \bibinfo {author} {\bibfnamefont {M.~B.}\ \bibnamefont
  {Plenio}},\ }\bibfield  {title} {\bibinfo {title} {Vibrations, quanta and
  biology},\ }\href {https://doi.org/10.1080/00405000.2013.829687} {\bibfield
  {journal} {\bibinfo  {journal} {Contemporary Physics}\ }\textbf {\bibinfo
  {volume} {54}},\ \bibinfo {pages} {181} (\bibinfo {year} {2013})}\BibitemShut
  {NoStop}%
\bibitem [{\citenamefont {Zerah~Harush}\ and\ \citenamefont
  {Dubi}(2021)}]{zerah_harush_photosynthetic_2021}%
  \BibitemOpen
  \bibfield  {author} {\bibinfo {author} {\bibfnamefont {E.}~\bibnamefont
  {Zerah~Harush}}\ and\ \bibinfo {author} {\bibfnamefont {Y.}~\bibnamefont
  {Dubi}},\ }\bibfield  {title} {\bibinfo {title} {Do photosynthetic complexes
  use quantum coherence to increase their efficiency? {Probably} not},\
  }\bibfield  {journal} {\bibinfo  {journal} {Sci Adv}\ }\textbf {\bibinfo
  {volume} {7}},\ \href {https://doi.org/10.1126/sciadv.abc4631}
  {10.1126/sciadv.abc4631} (\bibinfo {year} {2021})\BibitemShut {NoStop}%
\bibitem [{\citenamefont {Cao}\ \emph {et~al.}(2020)\citenamefont {Cao},
  \citenamefont {Cogdell}, \citenamefont {Coker}, \citenamefont {Duan},
  \citenamefont {Hauer}, \citenamefont {Kleinekathöfer}, \citenamefont
  {Jansen}, \citenamefont {Mančal}, \citenamefont {Miller}, \citenamefont
  {Ogilvie}, \citenamefont {Prokhorenko}, \citenamefont {Renger}, \citenamefont
  {Tan}, \citenamefont {Tempelaar}, \citenamefont {Thorwart}, \citenamefont
  {Thyrhaug}, \citenamefont {Westenhoff},\ and\ \citenamefont
  {Zigmantas}}]{cao_quantum_2020}%
  \BibitemOpen
  \bibfield  {author} {\bibinfo {author} {\bibfnamefont {J.}~\bibnamefont
  {Cao}}, \bibinfo {author} {\bibfnamefont {R.~J.}\ \bibnamefont {Cogdell}},
  \bibinfo {author} {\bibfnamefont {D.~F.}\ \bibnamefont {Coker}}, \bibinfo
  {author} {\bibfnamefont {H.-G.}\ \bibnamefont {Duan}}, \bibinfo {author}
  {\bibfnamefont {J.}~\bibnamefont {Hauer}}, \bibinfo {author} {\bibfnamefont
  {U.}~\bibnamefont {Kleinekathöfer}}, \bibinfo {author} {\bibfnamefont
  {T.~L.~C.}\ \bibnamefont {Jansen}}, \bibinfo {author} {\bibfnamefont
  {T.}~\bibnamefont {Mančal}}, \bibinfo {author} {\bibfnamefont {R.~J.~D.}\
  \bibnamefont {Miller}}, \bibinfo {author} {\bibfnamefont {J.~P.}\
  \bibnamefont {Ogilvie}}, \bibinfo {author} {\bibfnamefont {V.~I.}\
  \bibnamefont {Prokhorenko}}, \bibinfo {author} {\bibfnamefont
  {T.}~\bibnamefont {Renger}}, \bibinfo {author} {\bibfnamefont {H.-S.}\
  \bibnamefont {Tan}}, \bibinfo {author} {\bibfnamefont {R.}~\bibnamefont
  {Tempelaar}}, \bibinfo {author} {\bibfnamefont {M.}~\bibnamefont {Thorwart}},
  \bibinfo {author} {\bibfnamefont {E.}~\bibnamefont {Thyrhaug}}, \bibinfo
  {author} {\bibfnamefont {S.}~\bibnamefont {Westenhoff}},\ and\ \bibinfo
  {author} {\bibfnamefont {D.}~\bibnamefont {Zigmantas}},\ }\bibfield  {title}
  {\bibinfo {title} {Quantum biology revisited},\ }\href
  {https://doi.org/10.1126/sciadv.aaz4888} {\bibfield  {journal} {\bibinfo
  {journal} {Science Advances}\ }\textbf {\bibinfo {volume} {6}},\ \bibinfo
  {pages} {eaaz4888} (\bibinfo {year} {2020})}\BibitemShut {NoStop}%
\bibitem [{\citenamefont {Duan}\ \emph {et~al.}(2017)\citenamefont {Duan},
  \citenamefont {Prokhorenko}, \citenamefont {Cogdell}, \citenamefont {Ashraf},
  \citenamefont {Stevens}, \citenamefont {Thorwart},\ and\ \citenamefont
  {Miller}}]{duan_nature_2017}%
  \BibitemOpen
  \bibfield  {author} {\bibinfo {author} {\bibfnamefont {H.-G.}\ \bibnamefont
  {Duan}}, \bibinfo {author} {\bibfnamefont {V.~I.}\ \bibnamefont
  {Prokhorenko}}, \bibinfo {author} {\bibfnamefont {R.~J.}\ \bibnamefont
  {Cogdell}}, \bibinfo {author} {\bibfnamefont {K.}~\bibnamefont {Ashraf}},
  \bibinfo {author} {\bibfnamefont {A.~L.}\ \bibnamefont {Stevens}}, \bibinfo
  {author} {\bibfnamefont {M.}~\bibnamefont {Thorwart}},\ and\ \bibinfo
  {author} {\bibfnamefont {R.~J.~D.}\ \bibnamefont {Miller}},\ }\bibfield
  {title} {\bibinfo {title} {Nature does not rely on long-lived electronic
  quantum coherence for photosynthetic energy transfer},\ }\href
  {https://www-jstor-org.ezp-prod1.hul.harvard.edu/stable/26486911
  https://www.pnas.org/content/pnas/114/32/8493.full.pdf} {\bibfield  {journal}
  {\bibinfo  {journal} {Proceedings of the National Academy of Sciences of the
  United States of America}\ }\textbf {\bibinfo {volume} {114}},\ \bibinfo
  {pages} {8493} (\bibinfo {year} {2017})}\BibitemShut {NoStop}%
\bibitem [{\citenamefont {Mattioni}\ \emph {et~al.}(2021)\citenamefont
  {Mattioni}, \citenamefont {Caycedo-Soler}, \citenamefont {Huelga},\ and\
  \citenamefont {Plenio}}]{mattioni_design_2021}%
  \BibitemOpen
  \bibfield  {author} {\bibinfo {author} {\bibfnamefont {A.}~\bibnamefont
  {Mattioni}}, \bibinfo {author} {\bibfnamefont {F.}~\bibnamefont
  {Caycedo-Soler}}, \bibinfo {author} {\bibfnamefont {S.~F.}\ \bibnamefont
  {Huelga}},\ and\ \bibinfo {author} {\bibfnamefont {M.~B.}\ \bibnamefont
  {Plenio}},\ }\bibfield  {title} {\bibinfo {title} {Design {Principles} for
  {Long}-{Range} {Energy} {Transfer} at {Room} {Temperature}},\ }\href
  {https://doi.org/10.1103/PhysRevX.11.041003} {\bibfield  {journal} {\bibinfo
  {journal} {Physical Review X}\ }\textbf {\bibinfo {volume} {11}},\ \bibinfo
  {pages} {041003} (\bibinfo {year} {2021})}\BibitemShut {NoStop}%
\bibitem [{\citenamefont {Jang}\ and\ \citenamefont
  {Mennucci}(2018)}]{jang_delocalized_2018}%
  \BibitemOpen
  \bibfield  {author} {\bibinfo {author} {\bibfnamefont {S.~J.}\ \bibnamefont
  {Jang}}\ and\ \bibinfo {author} {\bibfnamefont {B.}~\bibnamefont
  {Mennucci}},\ }\bibfield  {title} {\bibinfo {title} {Delocalized excitons in
  natural light-harvesting complexes},\ }\href
  {https://doi.org/10.1103/RevModPhys.90.035003} {\bibfield  {journal}
  {\bibinfo  {journal} {Reviews of Modern Physics}\ }\textbf {\bibinfo {volume}
  {90}},\ \bibinfo {pages} {035003} (\bibinfo {year} {2018})}\BibitemShut
  {NoStop}%
\bibitem [{\citenamefont {Cao}\ and\ \citenamefont
  {Silbey}(2009)}]{cao_optimization_2009}%
  \BibitemOpen
  \bibfield  {author} {\bibinfo {author} {\bibfnamefont {J.}~\bibnamefont
  {Cao}}\ and\ \bibinfo {author} {\bibfnamefont {R.~J.}\ \bibnamefont
  {Silbey}},\ }\bibfield  {title} {\bibinfo {title} {Optimization of {Exciton}
  {Trapping} in {Energy} {Transfer} {Processes}},\ }\href
  {https://doi.org/10.1021/jp9032589} {\bibfield  {journal} {\bibinfo
  {journal} {The Journal of Physical Chemistry A}\ }\textbf {\bibinfo {volume}
  {113}},\ \bibinfo {pages} {13825} (\bibinfo {year} {2009})}\BibitemShut
  {NoStop}%
\bibitem [{\citenamefont {Chan}\ \emph {et~al.}(2018)\citenamefont {Chan},
  \citenamefont {Gamel}, \citenamefont {Fleming},\ and\ \citenamefont
  {Whaley}}]{chan_single-photon_2018}%
  \BibitemOpen
  \bibfield  {author} {\bibinfo {author} {\bibfnamefont {H.~C.~H.}\
  \bibnamefont {Chan}}, \bibinfo {author} {\bibfnamefont {O.~E.}\ \bibnamefont
  {Gamel}}, \bibinfo {author} {\bibfnamefont {G.~R.}\ \bibnamefont {Fleming}},\
  and\ \bibinfo {author} {\bibfnamefont {K.~B.}\ \bibnamefont {Whaley}},\
  }\bibfield  {title} {\bibinfo {title} {Single-photon absorption by single
  photosynthetic light-harvesting complexes},\ }\href
  {https://doi.org/10.1088/1361-6455/aa9c95} {\bibfield  {journal} {\bibinfo
  {journal} {Journal of Physics B: Atomic, Molecular and Optical Physics}\
  }\textbf {\bibinfo {volume} {51}},\ \bibinfo {pages} {054002} (\bibinfo
  {year} {2018})}\BibitemShut {NoStop}%
\bibitem [{\citenamefont
  {Lehmberg}(1970{\natexlab{a}})}]{lehmberg_radiation_1970}%
  \BibitemOpen
  \bibfield  {author} {\bibinfo {author} {\bibfnamefont {R.~H.}\ \bibnamefont
  {Lehmberg}},\ }\bibfield  {title} {\bibinfo {title} {Radiation from an {N}
  -{Atom} {System}. {II}. {Spontaneous} {Emission} from a {Pair} of {Atoms}},\
  }\href {https://doi.org/10.1103/PhysRevA.2.889} {\bibfield  {journal}
  {\bibinfo  {journal} {Physical Review A}\ }\textbf {\bibinfo {volume} {2}},\
  \bibinfo {pages} {889} (\bibinfo {year} {1970}{\natexlab{a}})}\BibitemShut
  {NoStop}%
\bibitem [{\citenamefont
  {Lehmberg}(1970{\natexlab{b}})}]{lehmberg_radiation_1970-1}%
  \BibitemOpen
  \bibfield  {author} {\bibinfo {author} {\bibfnamefont {R.~H.}\ \bibnamefont
  {Lehmberg}},\ }\bibfield  {title} {\bibinfo {title} {Radiation from an {N}
  -{Atom} {System}. {I}. {General} {Formalism}},\ }\href
  {https://doi.org/10.1103/PhysRevA.2.883} {\bibfield  {journal} {\bibinfo
  {journal} {Physical Review A}\ }\textbf {\bibinfo {volume} {2}},\ \bibinfo
  {pages} {883} (\bibinfo {year} {1970}{\natexlab{b}})}\BibitemShut {NoStop}%
\bibitem [{\citenamefont {Asenjo-Garcia}\ \emph {et~al.}(2017)\citenamefont
  {Asenjo-Garcia}, \citenamefont {Moreno-Cardoner}, \citenamefont {Albrecht},
  \citenamefont {Kimble},\ and\ \citenamefont
  {Chang}}]{asenjo-garcia_exponential_2017}%
  \BibitemOpen
  \bibfield  {author} {\bibinfo {author} {\bibfnamefont {A.}~\bibnamefont
  {Asenjo-Garcia}}, \bibinfo {author} {\bibfnamefont {M.}~\bibnamefont
  {Moreno-Cardoner}}, \bibinfo {author} {\bibfnamefont {A.}~\bibnamefont
  {Albrecht}}, \bibinfo {author} {\bibfnamefont {H.}~\bibnamefont {Kimble}},\
  and\ \bibinfo {author} {\bibfnamefont {D.}~\bibnamefont {Chang}},\ }\bibfield
   {title} {\bibinfo {title} {Exponential {Improvement} in {Photon} {Storage}
  {Fidelities} {Using} {Subradiance} and ``{Selective} {Radiance}'' in {Atomic}
  {Arrays}},\ }\href {https://doi.org/10.1103/PhysRevX.7.031024} {\bibfield
  {journal} {\bibinfo  {journal} {Physical Review X}\ }\textbf {\bibinfo
  {volume} {7}},\ \bibinfo {pages} {031024} (\bibinfo {year}
  {2017})}\BibitemShut {NoStop}%
\bibitem [{\citenamefont {Reitz}\ \emph {et~al.}(2022)\citenamefont {Reitz},
  \citenamefont {Sommer},\ and\ \citenamefont
  {Genes}}]{reitz_cooperative_2022}%
  \BibitemOpen
  \bibfield  {author} {\bibinfo {author} {\bibfnamefont {M.}~\bibnamefont
  {Reitz}}, \bibinfo {author} {\bibfnamefont {C.}~\bibnamefont {Sommer}},\ and\
  \bibinfo {author} {\bibfnamefont {C.}~\bibnamefont {Genes}},\ }\bibfield
  {title} {\bibinfo {title} {Cooperative {Quantum} {Phenomena} in
  {Light}-{Matter} {Platforms}},\ }\href
  {https://doi.org/10.1103/PRXQuantum.3.010201} {\bibfield  {journal} {\bibinfo
   {journal} {PRX Quantum}\ }\textbf {\bibinfo {volume} {3}},\ \bibinfo {pages}
  {010201} (\bibinfo {year} {2022})}\BibitemShut {NoStop}%
\bibitem [{\citenamefont {Peter}\ \emph
  {et~al.}(2024{\natexlab{a}})\citenamefont {Peter}, \citenamefont
  {Ostermann},\ and\ \citenamefont {Yelin}}]{peter_chirality_2024}%
  \BibitemOpen
  \bibfield  {author} {\bibinfo {author} {\bibfnamefont {J.~S.}\ \bibnamefont
  {Peter}}, \bibinfo {author} {\bibfnamefont {S.}~\bibnamefont {Ostermann}},\
  and\ \bibinfo {author} {\bibfnamefont {S.~F.}\ \bibnamefont {Yelin}},\
  }\bibfield  {title} {\bibinfo {title} {Chirality dependent photon transport
  and helical superradiance},\ }\href
  {https://doi.org/10.1103/PhysRevResearch.6.023200} {\bibfield  {journal}
  {\bibinfo  {journal} {Physical Review Research}\ }\textbf {\bibinfo {volume}
  {6}},\ \bibinfo {pages} {023200} (\bibinfo {year}
  {2024}{\natexlab{a}})}\BibitemShut {NoStop}%
\bibitem [{\citenamefont {Holzinger}\ \emph {et~al.}(2024)\citenamefont
  {Holzinger}, \citenamefont {Peter}, \citenamefont {Ostermann}, \citenamefont
  {Ritsch},\ and\ \citenamefont {Yelin}}]{holzinger_harnessing_2024}%
  \BibitemOpen
  \bibfield  {author} {\bibinfo {author} {\bibfnamefont {R.}~\bibnamefont
  {Holzinger}}, \bibinfo {author} {\bibfnamefont {J.~S.}\ \bibnamefont
  {Peter}}, \bibinfo {author} {\bibfnamefont {S.}~\bibnamefont {Ostermann}},
  \bibinfo {author} {\bibfnamefont {H.}~\bibnamefont {Ritsch}},\ and\ \bibinfo
  {author} {\bibfnamefont {S.}~\bibnamefont {Yelin}},\ }\bibfield  {title}
  {\bibinfo {title} {Harnessing quantum emitter rings for efficient energy
  transport and trapping},\ }\href {https://doi.org/10.1364/OPTICAQ.510021}
  {\bibfield  {journal} {\bibinfo  {journal} {Optica Quantum}\ }\textbf
  {\bibinfo {volume} {2}},\ \bibinfo {pages} {57} (\bibinfo {year}
  {2024})}\BibitemShut {NoStop}%
\bibitem [{\citenamefont {Peter}\ \emph
  {et~al.}(2024{\natexlab{b}})\citenamefont {Peter}, \citenamefont
  {Ostermann},\ and\ \citenamefont {Yelin}}]{peter_chirality-induced_2024}%
  \BibitemOpen
  \bibfield  {author} {\bibinfo {author} {\bibfnamefont {J.~S.}\ \bibnamefont
  {Peter}}, \bibinfo {author} {\bibfnamefont {S.}~\bibnamefont {Ostermann}},\
  and\ \bibinfo {author} {\bibfnamefont {S.~F.}\ \bibnamefont {Yelin}},\
  }\bibfield  {title} {\bibinfo {title} {Chirality-induced emergent spin-orbit
  coupling in topological atomic lattices},\ }\href
  {https://doi.org/10.1103/PhysRevA.109.043525} {\bibfield  {journal} {\bibinfo
   {journal} {Physical Review A}\ }\textbf {\bibinfo {volume} {109}},\ \bibinfo
  {pages} {043525} (\bibinfo {year} {2024}{\natexlab{b}})}\BibitemShut
  {NoStop}%
\bibitem [{\citenamefont {Dicke}(1954)}]{dicke_coherence_1954}%
  \BibitemOpen
  \bibfield  {author} {\bibinfo {author} {\bibfnamefont {R.~H.}\ \bibnamefont
  {Dicke}},\ }\bibfield  {title} {\bibinfo {title} {Coherence in {Spontaneous}
  {Radiation} {Processes}},\ }\href {https://doi.org/10.1103/PhysRev.93.99}
  {\bibfield  {journal} {\bibinfo  {journal} {Physical Review}\ }\textbf
  {\bibinfo {volume} {93}},\ \bibinfo {pages} {99} (\bibinfo {year}
  {1954})}\BibitemShut {NoStop}%
\bibitem [{\citenamefont {Gross}\ and\ \citenamefont
  {Haroche}(1982)}]{gross_superradiance_1982}%
  \BibitemOpen
  \bibfield  {author} {\bibinfo {author} {\bibfnamefont {M.}~\bibnamefont
  {Gross}}\ and\ \bibinfo {author} {\bibfnamefont {S.}~\bibnamefont
  {Haroche}},\ }\bibfield  {title} {\bibinfo {title} {Superradiance: {An} essay
  on the theory of collective spontaneous emission},\ }\href
  {https://doi.org/10.1016/0370-1573(82)90102-8} {\bibfield  {journal}
  {\bibinfo  {journal} {Physics Reports}\ }\textbf {\bibinfo {volume} {93}},\
  \bibinfo {pages} {301} (\bibinfo {year} {1982})}\BibitemShut {NoStop}%
\bibitem [{\citenamefont {Adolphs}\ and\ \citenamefont
  {Renger}(2006)}]{adolphs_how_2006}%
  \BibitemOpen
  \bibfield  {author} {\bibinfo {author} {\bibfnamefont {J.}~\bibnamefont
  {Adolphs}}\ and\ \bibinfo {author} {\bibfnamefont {T.}~\bibnamefont
  {Renger}},\ }\bibfield  {title} {\bibinfo {title} {How {Proteins} {Trigger}
  {Excitation} {Energy} {Transfer} in the {FMO} {Complex} of {Green} {Sulfur}
  {Bacteria}},\ }\href {https://doi.org/10.1529/biophysj.105.079483} {\bibfield
   {journal} {\bibinfo  {journal} {Biophysical Journal}\ }\textbf {\bibinfo
  {volume} {91}},\ \bibinfo {pages} {2778} (\bibinfo {year}
  {2006})}\BibitemShut {NoStop}%
\bibitem [{\citenamefont {Dorn}\ \emph {et~al.}(2011)\citenamefont {Dorn},
  \citenamefont {Strasfeld}, \citenamefont {Harris}, \citenamefont {Han},\ and\
  \citenamefont {Bawendi}}]{dorn_using_2011}%
  \BibitemOpen
  \bibfield  {author} {\bibinfo {author} {\bibfnamefont {A.}~\bibnamefont
  {Dorn}}, \bibinfo {author} {\bibfnamefont {D.~B.}\ \bibnamefont {Strasfeld}},
  \bibinfo {author} {\bibfnamefont {D.~K.}\ \bibnamefont {Harris}}, \bibinfo
  {author} {\bibfnamefont {H.-S.}\ \bibnamefont {Han}},\ and\ \bibinfo {author}
  {\bibfnamefont {M.~G.}\ \bibnamefont {Bawendi}},\ }\bibfield  {title}
  {\bibinfo {title} {Using {Nanowires} {To} {Extract} {Excitons} from a
  {Nanocrystal} {Solid}},\ }\href {https://doi.org/10.1021/nn203227t}
  {\bibfield  {journal} {\bibinfo  {journal} {ACS Nano}\ }\textbf {\bibinfo
  {volume} {5}},\ \bibinfo {pages} {9028} (\bibinfo {year} {2011})}\BibitemShut
  {NoStop}%
\bibitem [{\citenamefont {Lu}\ \emph {et~al.}(2009)\citenamefont {Lu},
  \citenamefont {Lingley}, \citenamefont {Asano}, \citenamefont {Harris},
  \citenamefont {Barwicz}, \citenamefont {Guha},\ and\ \citenamefont
  {Madhukar}}]{lu_photocurrent_2009}%
  \BibitemOpen
  \bibfield  {author} {\bibinfo {author} {\bibfnamefont {S.}~\bibnamefont
  {Lu}}, \bibinfo {author} {\bibfnamefont {Z.}~\bibnamefont {Lingley}},
  \bibinfo {author} {\bibfnamefont {T.}~\bibnamefont {Asano}}, \bibinfo
  {author} {\bibfnamefont {D.}~\bibnamefont {Harris}}, \bibinfo {author}
  {\bibfnamefont {T.}~\bibnamefont {Barwicz}}, \bibinfo {author} {\bibfnamefont
  {S.}~\bibnamefont {Guha}},\ and\ \bibinfo {author} {\bibfnamefont
  {A.}~\bibnamefont {Madhukar}},\ }\bibfield  {title} {\bibinfo {title}
  {Photocurrent {Induced} by {Nonradiative} {Energy} {Transfer} from
  {Nanocrystal} {Quantum} {Dots} to {Adjacent} {Silicon} {Nanowire}
  {Conducting} {Channels}: {Toward} a {New} {Solar} {Cell} {Paradigm}},\ }\href
  {https://doi.org/10.1021/nl903104k} {\bibfield  {journal} {\bibinfo
  {journal} {Nano Letters}\ }\textbf {\bibinfo {volume} {9}},\ \bibinfo {pages}
  {4548} (\bibinfo {year} {2009})}\BibitemShut {NoStop}%
\bibitem [{\citenamefont {Agranovich}\ \emph {et~al.}(2011)\citenamefont
  {Agranovich}, \citenamefont {Gartstein},\ and\ \citenamefont
  {Litinskaya}}]{agranovich_hybrid_2011}%
  \BibitemOpen
  \bibfield  {author} {\bibinfo {author} {\bibfnamefont {V.~M.}\ \bibnamefont
  {Agranovich}}, \bibinfo {author} {\bibfnamefont {Y.~N.}\ \bibnamefont
  {Gartstein}},\ and\ \bibinfo {author} {\bibfnamefont {M.}~\bibnamefont
  {Litinskaya}},\ }\bibfield  {title} {\bibinfo {title} {Hybrid {Resonant}
  {Organic}–{Inorganic} {Nanostructures} for {Optoelectronic}
  {Applications}},\ }\href {https://doi.org/10.1021/cr100156x} {\bibfield
  {journal} {\bibinfo  {journal} {Chemical Reviews}\ }\textbf {\bibinfo
  {volume} {111}},\ \bibinfo {pages} {5179} (\bibinfo {year}
  {2011})}\BibitemShut {NoStop}%
\bibitem [{\citenamefont {Lu}\ and\ \citenamefont
  {Madhukar}(2007)}]{lu_nonradiative_2007}%
  \BibitemOpen
  \bibfield  {author} {\bibinfo {author} {\bibfnamefont {S.}~\bibnamefont
  {Lu}}\ and\ \bibinfo {author} {\bibfnamefont {A.}~\bibnamefont {Madhukar}},\
  }\bibfield  {title} {\bibinfo {title} {Nonradiative {Resonant} {Excitation}
  {Transfer} from {Nanocrystal} {Quantum} {Dots} to {Adjacent} {Quantum}
  {Channels}},\ }\href {https://doi.org/10.1021/nl0719731} {\bibfield
  {journal} {\bibinfo  {journal} {Nano Letters}\ }\textbf {\bibinfo {volume}
  {7}},\ \bibinfo {pages} {3443} (\bibinfo {year} {2007})}\BibitemShut
  {NoStop}%
\bibitem [{\citenamefont {Caruso}\ \emph {et~al.}(2009)\citenamefont {Caruso},
  \citenamefont {Chin}, \citenamefont {Datta}, \citenamefont {Huelga},\ and\
  \citenamefont {Plenio}}]{caruso_highly_2009}%
  \BibitemOpen
  \bibfield  {author} {\bibinfo {author} {\bibfnamefont {F.}~\bibnamefont
  {Caruso}}, \bibinfo {author} {\bibfnamefont {A.~W.}\ \bibnamefont {Chin}},
  \bibinfo {author} {\bibfnamefont {A.}~\bibnamefont {Datta}}, \bibinfo
  {author} {\bibfnamefont {S.~F.}\ \bibnamefont {Huelga}},\ and\ \bibinfo
  {author} {\bibfnamefont {M.~B.}\ \bibnamefont {Plenio}},\ }\bibfield  {title}
  {\bibinfo {title} {Highly efficient energy excitation transfer in
  light-harvesting complexes: {The} fundamental role of noise-assisted
  transport},\ }\href {https://doi.org/10.1063/1.3223548} {\bibfield  {journal}
  {\bibinfo  {journal} {The Journal of Chemical Physics}\ }\textbf {\bibinfo
  {volume} {131}},\ \bibinfo {pages} {105106} (\bibinfo {year}
  {2009})}\BibitemShut {NoStop}%
\bibitem [{\citenamefont {Baumgratz}\ \emph {et~al.}(2014)\citenamefont
  {Baumgratz}, \citenamefont {Cramer},\ and\ \citenamefont
  {Plenio}}]{baumgratz_quantifying_2014}%
  \BibitemOpen
  \bibfield  {author} {\bibinfo {author} {\bibfnamefont {T.}~\bibnamefont
  {Baumgratz}}, \bibinfo {author} {\bibfnamefont {M.}~\bibnamefont {Cramer}},\
  and\ \bibinfo {author} {\bibfnamefont {M.}~\bibnamefont {Plenio}},\
  }\bibfield  {title} {\bibinfo {title} {Quantifying {Coherence}},\ }\href
  {https://doi.org/10.1103/PhysRevLett.113.140401} {\bibfield  {journal}
  {\bibinfo  {journal} {Physical Review Letters}\ }\textbf {\bibinfo {volume}
  {113}},\ \bibinfo {pages} {140401} (\bibinfo {year} {2014})}\BibitemShut
  {NoStop}%
\bibitem [{\citenamefont {Plenio}\ and\ \citenamefont
  {Virmani}(2007)}]{plenio_introduction_2007}%
  \BibitemOpen
  \bibfield  {author} {\bibinfo {author} {\bibfnamefont {M.}~\bibnamefont
  {Plenio}}\ and\ \bibinfo {author} {\bibfnamefont {S.}~\bibnamefont
  {Virmani}},\ }\bibfield  {title} {\bibinfo {title} {An introduction to
  entanglement measures},\ }\href {https://doi.org/10.26421/QIC7.1-2-1}
  {\bibfield  {journal} {\bibinfo  {journal} {Quantum Information and
  Computation}\ }\textbf {\bibinfo {volume} {7}},\ \bibinfo {pages} {1}
  (\bibinfo {year} {2007})}\BibitemShut {NoStop}%
\bibitem [{\citenamefont {Caruso}\ \emph {et~al.}(2010)\citenamefont {Caruso},
  \citenamefont {Chin}, \citenamefont {Datta}, \citenamefont {Huelga},\ and\
  \citenamefont {Plenio}}]{caruso_entanglement_2010}%
  \BibitemOpen
  \bibfield  {author} {\bibinfo {author} {\bibfnamefont {F.}~\bibnamefont
  {Caruso}}, \bibinfo {author} {\bibfnamefont {A.~W.}\ \bibnamefont {Chin}},
  \bibinfo {author} {\bibfnamefont {A.}~\bibnamefont {Datta}}, \bibinfo
  {author} {\bibfnamefont {S.~F.}\ \bibnamefont {Huelga}},\ and\ \bibinfo
  {author} {\bibfnamefont {M.~B.}\ \bibnamefont {Plenio}},\ }\bibfield  {title}
  {\bibinfo {title} {Entanglement and entangling power of the dynamics in
  light-harvesting complexes},\ }\href
  {https://doi.org/10.1103/PhysRevA.81.062346} {\bibfield  {journal} {\bibinfo
  {journal} {Physical Review A}\ }\textbf {\bibinfo {volume} {81}},\ \bibinfo
  {pages} {062346} (\bibinfo {year} {2010})}\BibitemShut {NoStop}%
\bibitem [{\citenamefont {Guo}\ \emph {et~al.}(2009)\citenamefont {Guo},
  \citenamefont {Salamo}, \citenamefont {Duchesne}, \citenamefont {Morandotti},
  \citenamefont {Volatier-Ravat}, \citenamefont {Aimez}, \citenamefont
  {Siviloglou},\ and\ \citenamefont {Christodoulides}}]{guo_observation_2009}%
  \BibitemOpen
  \bibfield  {author} {\bibinfo {author} {\bibfnamefont {A.}~\bibnamefont
  {Guo}}, \bibinfo {author} {\bibfnamefont {G.~J.}\ \bibnamefont {Salamo}},
  \bibinfo {author} {\bibfnamefont {D.}~\bibnamefont {Duchesne}}, \bibinfo
  {author} {\bibfnamefont {R.}~\bibnamefont {Morandotti}}, \bibinfo {author}
  {\bibfnamefont {M.}~\bibnamefont {Volatier-Ravat}}, \bibinfo {author}
  {\bibfnamefont {V.}~\bibnamefont {Aimez}}, \bibinfo {author} {\bibfnamefont
  {G.~A.}\ \bibnamefont {Siviloglou}},\ and\ \bibinfo {author} {\bibfnamefont
  {D.~N.}\ \bibnamefont {Christodoulides}},\ }\bibfield  {title} {\bibinfo
  {title} {Observation of {P} {T} -{Symmetry} {Breaking} in {Complex} {Optical}
  {Potentials}},\ }\href {https://doi.org/10.1103/PhysRevLett.103.093902}
  {\bibfield  {journal} {\bibinfo  {journal} {Physical Review Letters}\
  }\textbf {\bibinfo {volume} {103}},\ \bibinfo {pages} {093902} (\bibinfo
  {year} {2009})}\BibitemShut {NoStop}%
\bibitem [{\citenamefont {Joglekar}\ and\ \citenamefont
  {Harter}(2018)}]{joglekar_passive_2018}%
  \BibitemOpen
  \bibfield  {author} {\bibinfo {author} {\bibfnamefont {Y.~N.}\ \bibnamefont
  {Joglekar}}\ and\ \bibinfo {author} {\bibfnamefont {A.~K.}\ \bibnamefont
  {Harter}},\ }\bibfield  {title} {\bibinfo {title} {Passive
  parity-time-symmetry-breaking transitions without exceptional points in
  dissipative photonic systems},\ }\href {https://doi.org/10.1364/PRJ.6.000A51}
  {\bibfield  {journal} {\bibinfo  {journal} {Photonics Research}\ }\textbf
  {\bibinfo {volume} {6}},\ \bibinfo {pages} {A51} (\bibinfo {year}
  {2018})}\BibitemShut {NoStop}%
\bibitem [{\citenamefont {León-Montiel}\ \emph {et~al.}(2018)\citenamefont
  {León-Montiel}, \citenamefont {Quiroz-Juárez}, \citenamefont
  {Domínguez-Juárez}, \citenamefont {Quintero-Torres}, \citenamefont
  {Aragón}, \citenamefont {Harter},\ and\ \citenamefont
  {Joglekar}}]{leon-montiel_observation_2018}%
  \BibitemOpen
  \bibfield  {author} {\bibinfo {author} {\bibfnamefont {R.~D.~J.}\
  \bibnamefont {León-Montiel}}, \bibinfo {author} {\bibfnamefont {M.~A.}\
  \bibnamefont {Quiroz-Juárez}}, \bibinfo {author} {\bibfnamefont {J.~L.}\
  \bibnamefont {Domínguez-Juárez}}, \bibinfo {author} {\bibfnamefont
  {R.}~\bibnamefont {Quintero-Torres}}, \bibinfo {author} {\bibfnamefont
  {J.~L.}\ \bibnamefont {Aragón}}, \bibinfo {author} {\bibfnamefont {A.~K.}\
  \bibnamefont {Harter}},\ and\ \bibinfo {author} {\bibfnamefont {Y.~N.}\
  \bibnamefont {Joglekar}},\ }\bibfield  {title} {\bibinfo {title} {Observation
  of slowly decaying eigenmodes without exceptional points in {Floquet}
  dissipative synthetic circuits},\ }\href
  {https://doi.org/10.1038/s42005-018-0087-3} {\bibfield  {journal} {\bibinfo
  {journal} {Communications Physics}\ }\textbf {\bibinfo {volume} {1}},\
  \bibinfo {pages} {88} (\bibinfo {year} {2018})}\BibitemShut {NoStop}%
\bibitem [{\citenamefont {Syassen}\ \emph {et~al.}(2008)\citenamefont
  {Syassen}, \citenamefont {Bauer}, \citenamefont {Lettner}, \citenamefont
  {Volz}, \citenamefont {Dietze}, \citenamefont {García-Ripoll}, \citenamefont
  {Cirac}, \citenamefont {Rempe},\ and\ \citenamefont
  {Dürr}}]{syassen_strong_2008}%
  \BibitemOpen
  \bibfield  {author} {\bibinfo {author} {\bibfnamefont {N.}~\bibnamefont
  {Syassen}}, \bibinfo {author} {\bibfnamefont {D.~M.}\ \bibnamefont {Bauer}},
  \bibinfo {author} {\bibfnamefont {M.}~\bibnamefont {Lettner}}, \bibinfo
  {author} {\bibfnamefont {T.}~\bibnamefont {Volz}}, \bibinfo {author}
  {\bibfnamefont {D.}~\bibnamefont {Dietze}}, \bibinfo {author} {\bibfnamefont
  {J.~J.}\ \bibnamefont {García-Ripoll}}, \bibinfo {author} {\bibfnamefont
  {J.~I.}\ \bibnamefont {Cirac}}, \bibinfo {author} {\bibfnamefont
  {G.}~\bibnamefont {Rempe}},\ and\ \bibinfo {author} {\bibfnamefont
  {S.}~\bibnamefont {Dürr}},\ }\bibfield  {title} {\bibinfo {title} {Strong
  {Dissipation} {Inhibits} {Losses} and {Induces} {Correlations} in {Cold}
  {Molecular} {Gases}},\ }\href {https://doi.org/10.1126/science.1155309}
  {\bibfield  {journal} {\bibinfo  {journal} {Science}\ }\textbf {\bibinfo
  {volume} {320}},\ \bibinfo {pages} {1329} (\bibinfo {year}
  {2008})}\BibitemShut {NoStop}%
\bibitem [{\citenamefont {García-Ripoll}\ \emph {et~al.}(2009)\citenamefont
  {García-Ripoll}, \citenamefont {Dürr}, \citenamefont {Syassen},
  \citenamefont {Bauer}, \citenamefont {Lettner}, \citenamefont {Rempe},\ and\
  \citenamefont {Cirac}}]{garcia-ripoll_dissipation-induced_2009}%
  \BibitemOpen
  \bibfield  {author} {\bibinfo {author} {\bibfnamefont {J.~J.}\ \bibnamefont
  {García-Ripoll}}, \bibinfo {author} {\bibfnamefont {S.}~\bibnamefont
  {Dürr}}, \bibinfo {author} {\bibfnamefont {N.}~\bibnamefont {Syassen}},
  \bibinfo {author} {\bibfnamefont {D.~M.}\ \bibnamefont {Bauer}}, \bibinfo
  {author} {\bibfnamefont {M.}~\bibnamefont {Lettner}}, \bibinfo {author}
  {\bibfnamefont {G.}~\bibnamefont {Rempe}},\ and\ \bibinfo {author}
  {\bibfnamefont {J.~I.}\ \bibnamefont {Cirac}},\ }\bibfield  {title} {\bibinfo
  {title} {Dissipation-induced hard-core boson gas in an optical lattice},\
  }\href {https://doi.org/10.1088/1367-2630/11/1/013053} {\bibfield  {journal}
  {\bibinfo  {journal} {New Journal of Physics}\ }\textbf {\bibinfo {volume}
  {11}},\ \bibinfo {pages} {013053} (\bibinfo {year} {2009})}\BibitemShut
  {NoStop}%
\bibitem [{\citenamefont {Han}\ \emph {et~al.}(2009)\citenamefont {Han},
  \citenamefont {Chan}, \citenamefont {Yi}, \citenamefont {Daley},
  \citenamefont {Diehl}, \citenamefont {Zoller},\ and\ \citenamefont
  {Duan}}]{han_stabilization_2009}%
  \BibitemOpen
  \bibfield  {author} {\bibinfo {author} {\bibfnamefont {Y.-J.}\ \bibnamefont
  {Han}}, \bibinfo {author} {\bibfnamefont {Y.-H.}\ \bibnamefont {Chan}},
  \bibinfo {author} {\bibfnamefont {W.}~\bibnamefont {Yi}}, \bibinfo {author}
  {\bibfnamefont {A.~J.}\ \bibnamefont {Daley}}, \bibinfo {author}
  {\bibfnamefont {S.}~\bibnamefont {Diehl}}, \bibinfo {author} {\bibfnamefont
  {P.}~\bibnamefont {Zoller}},\ and\ \bibinfo {author} {\bibfnamefont {L.-M.}\
  \bibnamefont {Duan}},\ }\bibfield  {title} {\bibinfo {title} {Stabilization
  of the p -{Wave} {Superfluid} {State} in an {Optical} {Lattice}},\ }\href
  {https://doi.org/10.1103/PhysRevLett.103.070404} {\bibfield  {journal}
  {\bibinfo  {journal} {Physical Review Letters}\ }\textbf {\bibinfo {volume}
  {103}},\ \bibinfo {pages} {070404} (\bibinfo {year} {2009})}\BibitemShut
  {NoStop}%
\bibitem [{\citenamefont {Zhu}\ \emph {et~al.}(2014)\citenamefont {Zhu},
  \citenamefont {Gadway}, \citenamefont {Foss-Feig}, \citenamefont
  {Schachenmayer}, \citenamefont {Wall}, \citenamefont {Hazzard}, \citenamefont
  {Yan}, \citenamefont {Moses}, \citenamefont {Covey}, \citenamefont {Jin},
  \citenamefont {Ye}, \citenamefont {Holland},\ and\ \citenamefont
  {Rey}}]{zhu_suppressing_2014}%
  \BibitemOpen
  \bibfield  {author} {\bibinfo {author} {\bibfnamefont {B.}~\bibnamefont
  {Zhu}}, \bibinfo {author} {\bibfnamefont {B.}~\bibnamefont {Gadway}},
  \bibinfo {author} {\bibfnamefont {M.}~\bibnamefont {Foss-Feig}}, \bibinfo
  {author} {\bibfnamefont {J.}~\bibnamefont {Schachenmayer}}, \bibinfo {author}
  {\bibfnamefont {M.}~\bibnamefont {Wall}}, \bibinfo {author} {\bibfnamefont
  {K.}~\bibnamefont {Hazzard}}, \bibinfo {author} {\bibfnamefont
  {B.}~\bibnamefont {Yan}}, \bibinfo {author} {\bibfnamefont {S.}~\bibnamefont
  {Moses}}, \bibinfo {author} {\bibfnamefont {J.}~\bibnamefont {Covey}},
  \bibinfo {author} {\bibfnamefont {D.}~\bibnamefont {Jin}}, \bibinfo {author}
  {\bibfnamefont {J.}~\bibnamefont {Ye}}, \bibinfo {author} {\bibfnamefont
  {M.}~\bibnamefont {Holland}},\ and\ \bibinfo {author} {\bibfnamefont
  {A.}~\bibnamefont {Rey}},\ }\bibfield  {title} {\bibinfo {title} {Suppressing
  the {Loss} of {Ultracold} {Molecules} {Via} the {Continuous} {Quantum} {Zeno}
  {Effect}},\ }\href {https://doi.org/10.1103/PhysRevLett.112.070404}
  {\bibfield  {journal} {\bibinfo  {journal} {Physical Review Letters}\
  }\textbf {\bibinfo {volume} {112}},\ \bibinfo {pages} {070404} (\bibinfo
  {year} {2014})}\BibitemShut {NoStop}%
\bibitem [{\citenamefont {Peres}(1980)}]{peres_zeno_1980}%
  \BibitemOpen
  \bibfield  {author} {\bibinfo {author} {\bibfnamefont {A.}~\bibnamefont
  {Peres}},\ }\bibfield  {title} {\bibinfo {title} {Zeno paradox in quantum
  theory},\ }\href {https://doi.org/10.1119/1.12204} {\bibfield  {journal}
  {\bibinfo  {journal} {American Journal of Physics}\ }\textbf {\bibinfo
  {volume} {48}},\ \bibinfo {pages} {931} (\bibinfo {year} {1980})}\BibitemShut
  {NoStop}%
\bibitem [{\citenamefont {Haken}\ and\ \citenamefont
  {Strobl}(1973)}]{haken_exactly_1973}%
  \BibitemOpen
  \bibfield  {author} {\bibinfo {author} {\bibfnamefont {H.}~\bibnamefont
  {Haken}}\ and\ \bibinfo {author} {\bibfnamefont {G.}~\bibnamefont {Strobl}},\
  }\bibfield  {title} {\bibinfo {title} {An exactly solvable model for coherent
  and incoherent exciton motion},\ }\href {https://doi.org/10.1007/BF01399723}
  {\bibfield  {journal} {\bibinfo  {journal} {Zeitschrift für Physik A Hadrons
  and nuclei}\ }\textbf {\bibinfo {volume} {262}},\ \bibinfo {pages} {135}
  (\bibinfo {year} {1973})}\BibitemShut {NoStop}%
\bibitem [{\citenamefont {Čápek}(1993)}]{capek_hakenstroblreineker_1993}%
  \BibitemOpen
  \bibfield  {author} {\bibinfo {author} {\bibfnamefont {V.}~\bibnamefont
  {Čápek}},\ }\bibfield  {title} {\bibinfo {title}
  {Haken—{Strobl}—{Reineker} model: its limits of validity and a possible
  extension},\ }\href {https://doi.org/10.1016/0301-0104(93)85132-R} {\bibfield
   {journal} {\bibinfo  {journal} {Chemical Physics}\ }\textbf {\bibinfo
  {volume} {171}},\ \bibinfo {pages} {79} (\bibinfo {year} {1993})}\BibitemShut
  {NoStop}%
\bibitem [{\citenamefont {Leegwater}(1996)}]{leegwater_coherent_1996}%
  \BibitemOpen
  \bibfield  {author} {\bibinfo {author} {\bibfnamefont {J.~A.}\ \bibnamefont
  {Leegwater}},\ }\bibfield  {title} {\bibinfo {title} {Coherent versus
  {Incoherent} {Energy} {Transfer} and {Trapping} in {Photosynthetic} {Antenna}
  {Complexes}},\ }\href {https://doi.org/10.1021/jp961448i} {\bibfield
  {journal} {\bibinfo  {journal} {The Journal of Physical Chemistry}\ }\textbf
  {\bibinfo {volume} {100}},\ \bibinfo {pages} {14403} (\bibinfo {year}
  {1996})}\BibitemShut {NoStop}%
\bibitem [{\citenamefont {Brun}\ \emph {et~al.}(2003)\citenamefont {Brun},
  \citenamefont {Carteret},\ and\ \citenamefont
  {Ambainis}}]{brun_quantum_2003}%
  \BibitemOpen
  \bibfield  {author} {\bibinfo {author} {\bibfnamefont {T.~A.}\ \bibnamefont
  {Brun}}, \bibinfo {author} {\bibfnamefont {H.~A.}\ \bibnamefont {Carteret}},\
  and\ \bibinfo {author} {\bibfnamefont {A.}~\bibnamefont {Ambainis}},\
  }\bibfield  {title} {\bibinfo {title} {Quantum random walks with decoherent
  coins},\ }\href {https://doi.org/10.1103/PhysRevA.67.032304} {\bibfield
  {journal} {\bibinfo  {journal} {Physical Review A}\ }\textbf {\bibinfo
  {volume} {67}},\ \bibinfo {pages} {032304} (\bibinfo {year}
  {2003})}\BibitemShut {NoStop}%
\bibitem [{\citenamefont {Mirkovic}\ \emph {et~al.}(2017)\citenamefont
  {Mirkovic}, \citenamefont {Ostroumov}, \citenamefont {Anna}, \citenamefont
  {Van~Grondelle}, \citenamefont {{Govindjee}},\ and\ \citenamefont
  {Scholes}}]{mirkovic_light_2017}%
  \BibitemOpen
  \bibfield  {author} {\bibinfo {author} {\bibfnamefont {T.}~\bibnamefont
  {Mirkovic}}, \bibinfo {author} {\bibfnamefont {E.~E.}\ \bibnamefont
  {Ostroumov}}, \bibinfo {author} {\bibfnamefont {J.~M.}\ \bibnamefont {Anna}},
  \bibinfo {author} {\bibfnamefont {R.}~\bibnamefont {Van~Grondelle}}, \bibinfo
  {author} {\bibnamefont {{Govindjee}}},\ and\ \bibinfo {author} {\bibfnamefont
  {G.~D.}\ \bibnamefont {Scholes}},\ }\bibfield  {title} {\bibinfo {title}
  {Light {Absorption} and {Energy} {Transfer} in the {Antenna} {Complexes} of
  {Photosynthetic} {Organisms}},\ }\href
  {https://doi.org/10.1021/acs.chemrev.6b00002} {\bibfield  {journal} {\bibinfo
   {journal} {Chemical Reviews}\ }\textbf {\bibinfo {volume} {117}},\ \bibinfo
  {pages} {249} (\bibinfo {year} {2017})}\BibitemShut {NoStop}%
\bibitem [{\citenamefont {Breuer}\ and\ \citenamefont
  {Petruccione}(2010)}]{breuer_theory_2010}%
  \BibitemOpen
  \bibfield  {author} {\bibinfo {author} {\bibfnamefont {H.-P.}\ \bibnamefont
  {Breuer}}\ and\ \bibinfo {author} {\bibfnamefont {F.}~\bibnamefont
  {Petruccione}},\ }\href@noop {} {\emph {\bibinfo {title} {The theory of open
  quantum systems}}},\ \bibinfo {edition} {repr}\ ed.\ (\bibinfo  {publisher}
  {Clarendon Press},\ \bibinfo {address} {Oxford},\ \bibinfo {year}
  {2010})\BibitemShut {NoStop}%
\bibitem [{\citenamefont {Scully}\ and\ \citenamefont
  {Zubairy}(2008)}]{scully_quantum_2008}%
  \BibitemOpen
  \bibfield  {author} {\bibinfo {author} {\bibfnamefont {M.~O.}\ \bibnamefont
  {Scully}}\ and\ \bibinfo {author} {\bibfnamefont {M.~S.}\ \bibnamefont
  {Zubairy}},\ }\href@noop {} {\emph {\bibinfo {title} {Quantum optics}}},\
  \bibinfo {edition} {6th}\ ed.\ (\bibinfo  {publisher} {Cambridge Univ.
  Press},\ \bibinfo {address} {Cambridge},\ \bibinfo {year} {2008})\BibitemShut
  {NoStop}%
\bibitem [{\citenamefont {Shatokhin}\ \emph {et~al.}(2018)\citenamefont
  {Shatokhin}, \citenamefont {Walschaers}, \citenamefont {Schlawin},\ and\
  \citenamefont {Buchleitner}}]{shatokhin_coherence_2018}%
  \BibitemOpen
  \bibfield  {author} {\bibinfo {author} {\bibfnamefont {V.~N.}\ \bibnamefont
  {Shatokhin}}, \bibinfo {author} {\bibfnamefont {M.}~\bibnamefont
  {Walschaers}}, \bibinfo {author} {\bibfnamefont {F.}~\bibnamefont
  {Schlawin}},\ and\ \bibinfo {author} {\bibfnamefont {A.}~\bibnamefont
  {Buchleitner}},\ }\bibfield  {title} {\bibinfo {title} {Coherence turned on
  by incoherent light},\ }\href {https://doi.org/10.1088/1367-2630/aaf08f}
  {\bibfield  {journal} {\bibinfo  {journal} {New Journal of Physics}\ }\textbf
  {\bibinfo {volume} {20}},\ \bibinfo {pages} {113040} (\bibinfo {year}
  {2018})}\BibitemShut {NoStop}%
\bibitem [{\citenamefont {Noschese}\ \emph {et~al.}(2013)\citenamefont
  {Noschese}, \citenamefont {Pasquini},\ and\ \citenamefont
  {Reichel}}]{noschese_tridiagonal_2013}%
  \BibitemOpen
  \bibfield  {author} {\bibinfo {author} {\bibfnamefont {S.}~\bibnamefont
  {Noschese}}, \bibinfo {author} {\bibfnamefont {L.}~\bibnamefont {Pasquini}},\
  and\ \bibinfo {author} {\bibfnamefont {L.}~\bibnamefont {Reichel}},\
  }\bibfield  {title} {\bibinfo {title} {Tridiagonal {Toeplitz} matrices:
  properties and novel applications: {TRIDIAGONAL} {TOEPLITZ} {MATRICES}},\
  }\href {https://doi.org/10.1002/nla.1811} {\bibfield  {journal} {\bibinfo
  {journal} {Numerical Linear Algebra with Applications}\ }\textbf {\bibinfo
  {volume} {20}},\ \bibinfo {pages} {302} (\bibinfo {year} {2013})}\BibitemShut
  {NoStop}%
\bibitem [{\citenamefont {Kulkarni}\ \emph {et~al.}(1999)\citenamefont
  {Kulkarni}, \citenamefont {Schmidt},\ and\ \citenamefont
  {Tsui}}]{kulkarni_eigenvalues_1999}%
  \BibitemOpen
  \bibfield  {author} {\bibinfo {author} {\bibfnamefont {D.}~\bibnamefont
  {Kulkarni}}, \bibinfo {author} {\bibfnamefont {D.}~\bibnamefont {Schmidt}},\
  and\ \bibinfo {author} {\bibfnamefont {S.-K.}\ \bibnamefont {Tsui}},\
  }\bibfield  {title} {\bibinfo {title} {Eigenvalues of tridiagonal
  pseudo-{Toeplitz} matrices},\ }\href
  {https://doi.org/10.1016/S0024-3795(99)00114-7} {\bibfield  {journal}
  {\bibinfo  {journal} {Linear Algebra and its Applications}\ }\textbf
  {\bibinfo {volume} {297}},\ \bibinfo {pages} {63} (\bibinfo {year}
  {1999})}\BibitemShut {NoStop}%
\end{thebibliography}%

\end{document}